\documentclass{aa}  
\usepackage{longtable}
\usepackage{caption}
\usepackage{subcaption}
\usepackage{amsmath}
\usepackage{graphicx}
\usepackage{txfonts}
\usepackage[fleqn]{amsmath}	
\usepackage{amssymb}	
\usepackage{mathtools}
\usepackage[breaklinks=true]{hyperref}
\hypersetup{
  colorlinks=true,   
  citecolor=blue,    
  urlcolor=blue,     
  linkcolor=blue,     
}
\bibpunct{(}{)}{;}{a}{}{,} 
\usepackage{lscape}

\DeclareRobustCommand{\ion}[2]{%
\relax\ifmmode
\ifx\testbx\f@series
{\mathbf{#1\,\mathsc{#2}}}\else
{\mathrm{#1\,\mathsc{#2}}}\fi
\else\textup{#1\,{\mdseries\textsc{#2}}}%
\fi}

   \newcommand{\matplotlib}{\texttt{matplotlib}\xspace}
   \newcommand{\topcat}{\texttt{TOPCAT}\xspace}

\begin{document}

    \title{The near-infrared SED of blue quasars: what drives the evolution of the dusty torus?}

   \author{Bartolomeo Trefoloni\inst{1,2}
   \thanks{\email{bartolomeo.trefoloni@unifi.it}},
    Roberto~Gilli\inst{3},
    Elisabeta~Lusso\inst{1,2}, 
    Alessandro~Marconi\inst{1,2},
    Giovanni~Mazzolari\inst{3,4},
    Emanuele~Nardini\inst{2},
    Guido~Risaliti\inst{1,2},
    Matilde~Signorini\inst{2,5}
	      }
          
\institute{
$^{1}$Dipartimento di Fisica e Astronomia, Universit\`a di Firenze, via G. Sansone 1, 50019 Sesto Fiorentino, Firenze, Italy\\
$^{2}$INAF -- Osservatorio Astrofisico di Arcetri, Largo Enrico Fermi 5, I-50125 Firenze, Italy\\
$^{3}$INAF -- Osservatorio di Astrofisica e Scienza dello Spazio di Bologna, via Gobetti 93/3, I-40129 Bologna, Italy\\
$^{4}$Dipartimento di Fisica e Astronomia, Università di Bologna, Via Gobetti 93/2, I-40129 Bologna, Italy\\
$^{5}$Dipartimento di Matematica e Fisica, Univeristà di Roma 3, Via della Vasca Navale, 84, 00146 Roma RM\\
\\
}

\titlerunning{The near-infrared SED of blue quasars: what drives the evolution of the dusty torus?}
\authorrunning{B. Trefoloni et al.}

\abstract{A fundamental ingredient in the unified model of active galactic nuclei (AGN) is the obscuring torus, whose innermost, hottest region dominates the near infrared (NIR) emission. Characterising the change in the torus properties and its interplay with the main AGN emission is key for our understanding of AGN physics, evolution and classification. Its covering factor ($CF$) is largely responsible for the classification of AGN on the basis of the detection of broad emission lines. It is still not clear whether the torus properties evolve over time and how they relate with the accretion parameters of the nucleus. In this work, we aim at investigating the evolution of the NIR properties with the redshift ($z$) and the bolometric luminosity ($L_{\rm bol}$) of the AGN. To this end, we assembled a large dataset of $\sim$36,000 Type 1 AGN between $0.5<z<2.9$ and $45.0<\log(L_{\rm bol} / (\rm erg / s))<48.0$ with UV, optical and near-infrared photometry. We produced average spectral energy distributions (SED) in different bins of the $z-L_{\rm bol}$ parameter space to estimate how the NIR SED evolves according to these parameters. We find that the NIR luminosity decreases for increasing $L_{\rm bol}$ at any redshift. At the same, time the shape of the NIR SED in our sample is consistent with a non-evolution with $z$. As a consequence, all the explored proxies for the $CF$ exhibit significant anti-correlations with $L_{\rm bol}$, but not with $z$. Additionally, the $CF$ also shows a shallower anti-correlation with the Eddington ratio ($\lambda_{\rm Edd}$), yet current systematic uncertainties, as well as the limited dynamical range, do not allow us to precisely constrain the role of the Eddington ratio. Lastly, we derived the covering factor from the ratio between the NIR and optical luminosity and we employed it to set a lower limit for the X-ray obscuration at different redshifts.}


   \maketitle
%

\section{Introduction}

Supermassive black holes (SMBH) efficiently accreting gas and dust are known as active galactic nuclei (AGN). The spectral energy distribution (SED) of these sources covers the whole electromagnetic spectrum, with two noticeable peaks: the ``big blue bump'' (BBB, \citealt{shields1978thermal}) in the optical--UV, and an infrared bump around 10 $\mu m$ (e.g. \citealt{edelson1986spectral, barvainis1987}). The former is thought to be produced by the accretion disc powering the AGN, while the latter arises because of some colder gaseous and dusty material re-emitting the the main AGN continuum in the infrared.

In the current standard picture, the so-called unified model (\citealt{antonucci1993unified, urry1995unified}), the observational diversity of AGN in terms of spectral properties is explained as the result of different lines of sight to the central engine. A key ingredient of the unified model is the obscuring torus, a dusty and gaseous structure on pc scales, surrounding the accretion disc. In this framework, an unobscured (type 1) AGN is observed when the line of sight does not cross the obscuring material, while the inner engine is hidden by the toroidal obscurer in case of an obscured (type 2) AGN.
This observational classification can also be interpreted in terms of escape probability of the light through the spatial distribution of clumpy elements forming the torus, thus allowing for a less strict division between Type 1 and Type 2 AGN (see e.g. \citealt{almeida2011testing, mateos2016x}). An outer tilted torus on scales $\sim$100 pc (\citealt{mcleod1995near, maiolino1995low}) can also be included to account for intermediate AGN types. 

A critical parameter defining the properties of the torus is the covering factor ($CF$), which is generally defined as the fraction of the whole solid angle covered by an obscuring structure ($\Omega / 4\pi$, e.g. \citealt{hamann1993geometry}). Although the definition of the covering factor is purely geometrical, on a practical level there are different routes to estimate this parameter, taking advantage of several techniques across different wave-bands. 

In the X-ray surveys, the fraction of obscured sources (generally those with $\log (N_{\rm H} / cm^{-2} )>22$) is used as a proxy for the mean covering factor at a given luminosity, given a sample with a high enough completeness level and under the hypothesis of randomly oriented AGN in the sky (e.g., \citealt{ueda2003cosmological, ueda2014toward, la2005hellas2xmm, hasinger2008absorption, burlon2011three, malizia2012integral, merloni2014incidence}). A similar argument has been used to estimate the $CF$ as the ratio between narrow line (Type 2) and broad line (Type 1 AGN), based on the optical spectrum (\citealt{simpson2005luminosity, toba2014luminosity}). 

In addition, by considering that the intrinsic UV luminosity of the AGN is reprocessed by the gaseous and dusty torus and re-emitted in the infrared (e.g., \citealt{maiolino2007dust, treister2008measuring, gandhi2009resolving}), the covering factor can thus be derived as the ratio between the reprocessed infrared luminosity and the intrinsic AGN emission. Yet, this ratio does not take into account the complicated anatomy of the covering material (\citealt{granato1994thick}). Early analytical models employed a simplistic continuous toroidal screen (\citealt{pier1992infrared}), but the need for a more irregular gas and dust distribution was already clear. The strong nuclear radiation field is indeed expected to break a continuous torus into smaller clumps (\citealt{krolik1988molecular}) whose turbulent motion helps in stabilising a geometrically thick structure (\citealt{beckert2004dynamical}). Deriving the $CF$ is far less straightforward for a non-homogeneous (i.e. clumpy) distribution of the clouds composing the torus (\citealt{nenkova2008agn1, nenkova2008agn2, lusso2013obscured, netzer2015revisiting}), and detailed radiative transfer modelling \citep{stalevski2016dust} showed that the bare infrared-to-optical luminosity ratio systematically underestimates low $CF$s and overestimates high $CF$s in case of type 1 AGN.

Critical pieces of information about the torus properties have also been gathered by successfully applying torus models to reproduce observations of low redshift Seyferts and luminous AGN both in the infrared (e.g. \citealt{mor2009dusty, honig2010dusty, alonso2011torus, alonso2021galaxy, almeida2011testing, feltre2012smooth, zhuang2018infrared, ichikawa2019bat, gonzalez2019exploring, garcia2019torus, garcia2022torus}) and in the X-rays (see e.g. \citealt{ricci2023ray}). In the latter case, both the iron $K\alpha$ emission line and the $\sim$\,30 keV Compton hump have been used to constrain the geometry of the circumnuclear material (\citealt{ricci2014iron, gandhi2015dust}).

In a dynamical picture, the properties of the torus are expected to vary according to the accretion parameters of the AGN (i.e. the bolometric luminosity $L_{\rm bol}$, the black hole mass $M_{\rm BH}$ and the Eddington ratio $\lambda_{\rm Edd}$). A simple evolutionary scenario predicts that its covering factor should decrease with increasing luminosity, i.e. the idea of the ``receding torus'' (\citealt{lawrence1991relative}), or a ``luminosity-dependent unified model'' (\citealt{ueda2003cosmological}). In this picture, the radiation field produced in the innermost regions of the AGN is capable of directly shaping the morphology of the gaseous and dusty surrounding material. At the same time, efficient accretion and high luminosity are expected to be conducive conditions for the launch of powerful accretion-disc winds, directly driven by the nuclear activity (e.g. \citealt{proga2005, zubovas2013bal, nardini2015black, king2015powerful, tombesi2015wind, nardini2019}), which could be equally able to shape the circumnuclear environment in episodic events rather than on galactic timescales. A dependence on the accretion rate is also supported by recent studies (see e.g. \citealt{ricci2023bass} and references therein) arguing that the actual driver of the covering factor decrease might be the Eddington ratio rather than the bare bolometric luminosity, with the former being equivalent to an effective (mass-normalized) luminosity.

Although the properties of the torus, dominating the infrared SED of AGN, are expected to vary with the luminosity and/or the Eddington ratio, its possible evolution with redshift has not yet found credible evidence. From a cosmological standpoint, there is growing evidence that the UV and optical SED properties in quasars do not appreciably evolve with redshift. Luminous blue AGN shining at high redshift are indeed remarkably similar to those observed locally, both in terms of continuum (e.g. \citealt{kuhn2001search, mortlock2011luminous, hao2013, shen2019gemini, yang2021probing, trefoloni2024quasars}) and line (e.g. \citealt{croom2002correlation, fan2004survey, nagao2006, stepney2023no}) properties. These findings are echoed in the infrared where several works showed that the average SEDs made of samples over wide redshift intervals, up to $z\sim4$, exhibit quite similar shapes also in the NIR and MIR ranges (e.g. \citealt{elvis1994atlas, richards2006spectral, krawczyk2013mean}, but see also \citealt{bianchini2019broadband} for recent updates from a larger MIR sample). 
This evidence is confirmed also by other sparse samples at $z\gtrsim5$, which exhibit NIR properties closely resembling those of the bulk of the quasar population at lower redshifts (\citealt{leipski2014spectral}), although with the noticeable exception of hot dust deficient quasars (e.g. \citealt{jiang2010dust, hao2011hot}).
Yet, a systematic investigation of the NIR properties, and their interplay with the main optical and UV continuum, separately analysing the effects of luminosity and redshift over wide intervals, has not yet been undertaken for a large sample of blue unobscured AGN.

In this work, our aim is to investigate separately the possible evolution of the NIR SED in a sample of blue Type 1 AGN with the accretion parameters ($L_{\rm bol}$, $\lambda_{\rm Edd}$) and the redshift, taking advantage of a large complete dataset purposefully assembled.
In particular, we employed this large sample to address the following issues: i) investigate the evolution with $L_{\rm bol}$ and $z$ of the NIR SED in blue quasars; ii) investigate how the covering factor in quasars evolves separately with $z$ and with $L_{\rm bol}$; iii) explore the tentative anti-correlation between $CF$ and $\lambda_{\rm Edd}$.

The paper is structured as follows. A brief summary about the selection of the quasar sample is provided in Section \ref{sec:dataset}, whilst the analysis performed on the sample SED is described in Section \ref{sec:methods}. Results are discussed in Section \ref{sec:results} and conclusions are drawn in Section \ref{sec:discussion}. 

Throughout this paper we adopt a flat $\Lambda$CDM cosmology with $H_0 = 70$ km s$^{-1}$ Mpc$^{-1}$, $\Omega_{\Lambda}$ = 0.7, and $\Omega_{m}$ = 0.3. Correlations are considered significant when yielding p-values lower than 0.01, corresponding to $\sim$2.6 $\sigma$ in the case of Normal approximation, under the null-hypothesis of non correlation.

\section{The data-set}
\label{sec:dataset}

With the aim of building a dataset covering the rest-frame emission from the UV to the IR for a large sample, we gathered and cross-matched photometric data from different surveys. Here we briefly highlight the building blocks of the final sample.

The starting sample is the Sloan Digital Sky Survey (SDSS) DR7 (\citealt{shen2011}), which samples the rest-frame optical and UV wavelengths, where the AGN SED is dominated by the disc emission. This contains $\sim$106,000 quasars with both photometric and spectral data in the five SDSS bands (\citealt{fukugita1996sloan}) brighter than $M_i<-22.0$. Although newer releases have been produced, we opted for the SDSS DR7 catalogue as it offers a subsample of roughly 50,000 objects which abide by a uniform selection criterion, defined using the final quasar target selection algorithm described in \citet{richards2002}, identified by the flag UNIFORM=1 in the catalogue. This avoids the uneven sampling along the redshift axis, and the selection criteria optimised to map the large-scale structure traced by the Ly-$\alpha$ forest, that instead characterises the later BOSS (\citealt{dawson2012baryon}) and EBOSS (\citealt{dawson2016sdss}) surveys. At the same time, the SDSS selection algorithm claims a completeness close to $\sim 90\%$ below $m_i<19.1$ at $z<2.9$ (see Sec 4.1 in \citealt{richards2002} for the related caveats), and this protects this work from incompleteness biases. With the aim of further increasing the quality of the sample we eventually opted for a magnitude cut at $m_i\leq 19.0$. Data in the $u$ and $z$ filters were corrected for the reported shifts in their zeropoints\footnote{https://www.sdss4.org/dr17/algorithms/fluxcal/}.

The near-infrared band ($\sim$ 1--10 $\mu$m) is crucial when investigating the emission of the innermost hot region of the torus, which dominates the SED of luminous AGN in this band. At these wavelengths the AllWISE catalogue (\citealt{cutri2021vizier}) is the largest dataset available. This catalogue was produced from the collection of the Wide Infrared Survey Explorer (WISE, \citealt{wright2010wide}) observations. We cross-matched this sample with the SDSS one adopting an optimal 2$^{\prime\prime}$ matching radius (see Sec. 2.2 in \citealt{krawczyk2013mean}), which gives $\sim 97\%$ of matches.
For a typical quasar SED, the WISE filters have effective wavelengths 3.36 $\mu$m ($W1$), 4.61 $\mu$m ($W2$), 11.82 $\mu$m ($W3$) and 22.13 $\mu$m ($W4$). To improve the quality of the NIR data, we also required at least SNR=2 in W3 band, a requirement easily met in both W1 and W2. More details about the reliability of the WISE photometry and our NIR quality cuts can be found in Appendix \ref{app:wise_phot}.
Although the features of the AGN SED we are interested in are already well covered 
by these two surveys, we complemented the sample with other near-infrared and UV data. In particular, we added the UV points from the Galaxy Evolution Explorer (GALEX, \citealt{martin2005}). The observed-frame reference wavelengths of the GALEX filters are 1528 \AA~(FUV) and 2271 \AA~(NUV), yielding respectively 38,000 and 45,000 detections which helped to probe the peak of the disc emission especially in the low redshift regime. 
In the near-infrared (NIR) band the main contributions to the SED come from the low-energy tail of the disc and the stellar emission from the host galaxy, which we covered using NIR photometry from both the UKIRT (United Kingdom Infrared Telescope) Infrared
Deep Sky Survey (UKIDSS; \citealt{lawrence2007}) and the Two Micron All Sky Survey (2MASS; \citealt{skrutskie2006two}). These two supplementary datasets provided photometric data for roughly 11,300 and 10,000 sources in the $YJHK$ and $JHK$ bandpasses, respectively.

All the available photometry was also corrected for Galactic extinction adopting the \cite{fitzpatrick1999correcting} extinction curve and the extinction values from \citet{schlafly2011measuring}.

\subsection{Accounting for selection effects}
Since one of the goals of this work is to estimate if any evolution with $z$ of $CF$ exists, starting from the cross-match of flux-limited surveys, it is extremely important to take into account to what extent selection effects can affect our results. These can come basically in two forms: incompleteness and inclination.
The first effect is produced whenever the optical detection is not matched by the IR counterpart in a substantial fraction of objects. This implies that only the most IR-bright objects are selected because of the IR flux sensitivity, thus biasing the $CF$ towards higher-than-average values. In our case this issue is not critical, as the IR-to-optical cross-match fraction ($\sim 97\%$) guarantees a high degree of completeness with respect to the optical survey, which also claims to be $\sim 90\%$ complete.

The second effect has a subtler impact on the observations. We assume that the optical/UV emission comes from an accretion disc, while the IR radiation is isotropically emitted by the torus (see e.g. \citealt{treister2006evolution} and references therein). In flux-limited surveys, at higher redshifts increasingly fainter objects fall below the flux limit of the survey. Assuming that the intrinsic luminosity is diminished because of the projection effect as $L_{\rm bol,obs} = L_{\rm bol,int} \, \cos \theta$, with $\theta$ being the angle between the line of sight and the axis of the accretion disc, the same observed luminosity can be obtained by intrinsically brighter sources observed at higher inclination or vice-versa. Even assuming that all the quasars in the Universe had the same intrinsic luminosity (which is obviously not true) and random inclinations of the line of sight, the combined effect of the flux limit and the increasing distance with $z$ would imply the selection of preferentially face-on objects. This, in turn translates into a relatively higher contribution of the optical/UV with respect to the IR (i.e. lower $CF$) for increasing $z$.

Such an effect, would produce an anti-correlation between the $CF$ and $z$ at a given intrinsic luminosity, but in the following we will demonstrate that, within the ranges of parameters explored in this work, it does not produce any statistically appreciable trend. Another, likely overly fine-tuned  possibility is that the effect of increasingly face-on objects combines with intrinsically higher $CF$ so that the two effects cancel out. Similarly, the anti-correlation observed in the data-set between between $CF$ and $L_{\rm bol}$ at constant $z$ (see Sec. \ref{sec:nir_evolution}, \ref{sec:lbol_correlation}), under the reasonable assumption of random inclinations of the line of sight at the same redshift, cannot be ascribed to inclination effects. Therefore this anti-correlation between $CF$ and $L_{\rm bol}$ that we observe is likely a genuine feature rather than one induced by observational effects.

To quantify the impact of the inclination angle at increasing redshifts on the observed $CF$ we took advantage of a simulation. The basic idea here is to simulate a population of quasars whose intrinsic bolometric luminosity and covering factor distribution are known and see how the effect of inclination and flux limit propagate in the observed $CF$ in the parameter space defined by $L_{\rm bol}$ and $z$. To this end, we produced a mock sample of 100,000 objects uniformly distributed in the parameter space. The limits of the $z$ and $\log(L_{\rm bol})$ distributions overlap with those spanned by the actual sample, i.e. $z$=[0.1,2.9] and $\log(L_{\rm bol})$=45.0--48.5 (the upper limit of the luminosity was chosen to be slightly higher so as to include very luminous, yet inclined objects). We assumed a normal intrinsic $CF$ distribution with mean value $\langle CF \rangle =0.5$ and standard deviation $\sigma_{CF}=0.05$, with the underlying assumption that there is no physical connection between the torus $CF$ and the AGN luminosity.
For each object we draw a random value of $\cos\theta$ from a uniform distribution between [$\cos 0^{\circ},\cos 90^{\circ}$], representing random line of sights to the AGN of the sample, and derived the projected bolometric luminosity accordingly. This allowed us to produce the observed $CF_{\rm obs} = CF / \cos\theta $.

Additionally, we produced the simulated photometry by assuming a 3000 \AA\, luminosity $L_{3000\AA}$=$L_{\rm bol}$/5.15 from the bolometric corrections from \citet{richards2006spectral} and matching a quasar template (\citealt{vandenberk2001}) to  $L_{3000\AA}$. Finally, we discarded all the objects whose $i$ photometry fell above $m_i \geq 19.0$ (the magnitude limit chosen for this work, see Sec.\ref{sec:dataset}) and whose inclination angle was above 65$^{\circ}$,  assuming an infinitely optically thick torus. Above this level of inclination, the broad lines would not be detected and the optical/UV SED would not exhibit the blue colours of unobscured quasars and would not be consequently included in the survey.

\begin{figure*}[h!]
\begin{subfigure}{0.5\textwidth}
\includegraphics[width=0.95\linewidth]{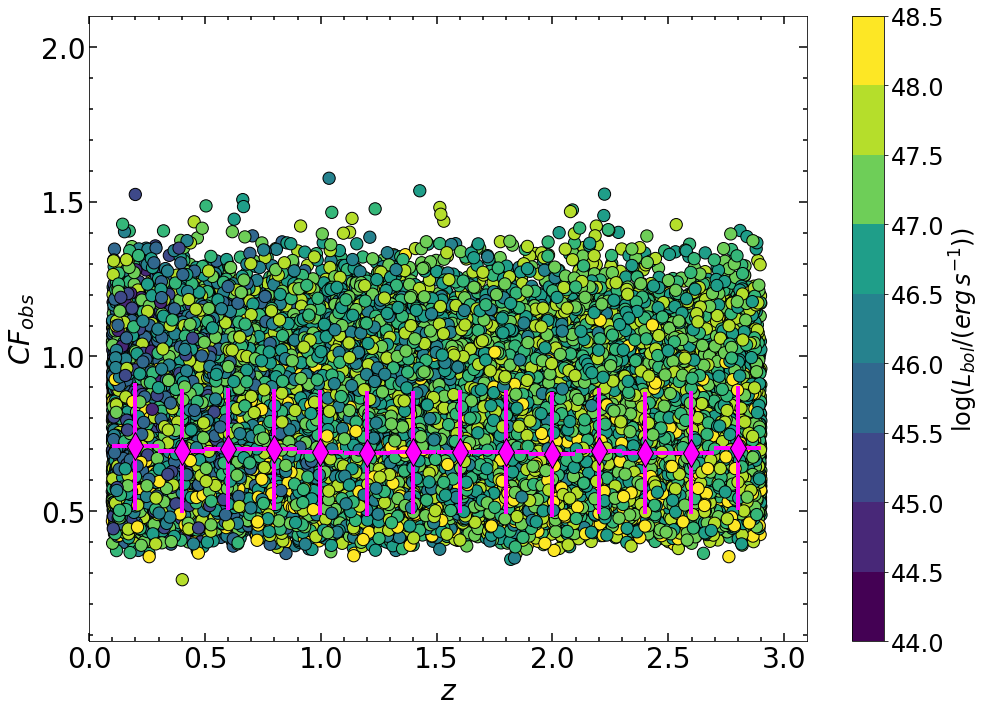}
\end{subfigure}
\hspace{0.02\textwidth}
\begin{subfigure}{0.5\textwidth}
\includegraphics[width=0.95\linewidth]{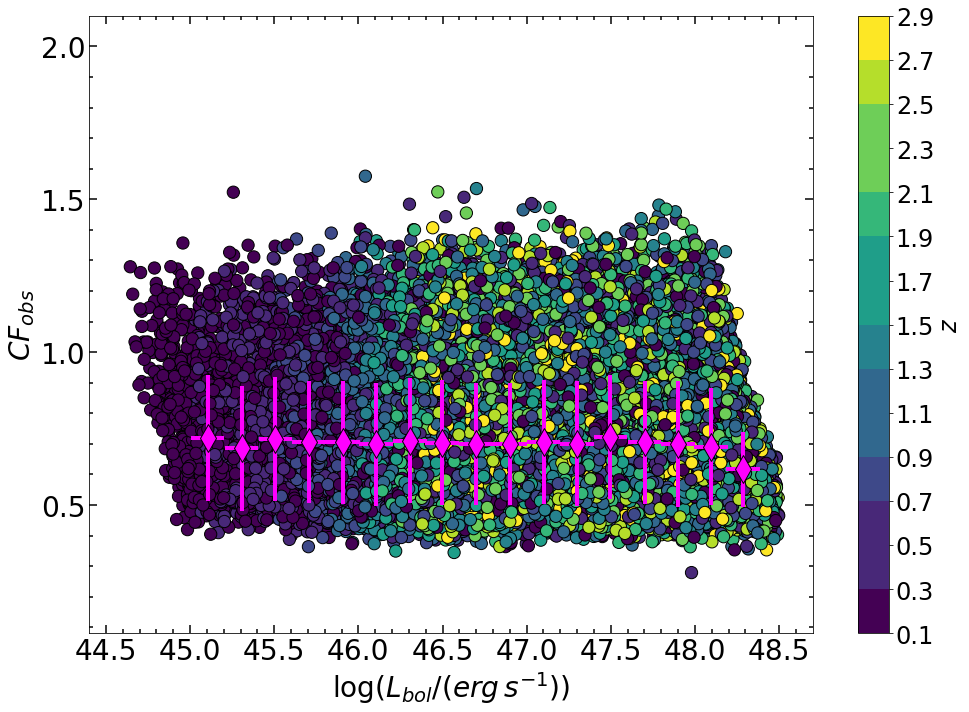}
\end{subfigure}
\caption{\textit{Left:} $z$--$CF$ (left) and $\log(L_{\rm bol})$--$CF$ (right) section of the simulated parameter space. The $CF$ and $\log(L_{\rm bol})$ values presented here are the observed (i.e. projected) quantities. Magenta diamonds represent the median $CF$ values in bins 0.2-dex wide. The $y$-axis uncertainties represent the standard error on the mean $CF$ in the bin. The third axis is colour-coded.}
    
\label{fig:mc_sections}
\end{figure*}

In Fig. \ref{fig:mc_sections} we show the result of this procedure. There are several details to be noted here: the first and foremost is that the combined effects of inclination and flux limit are not capable of producing any appreciable trend in the two sections of the parameter space. Secondly, we note that the average observed covering factor $CF_{\rm obs}$ is $\sim$0.7, while the intrinsic one set in the simulation was $CF$=0.5. This can be explained as the effect of the average  inclination by which $CF_{\rm obs}=CF / \langle \cos \, \theta \rangle \simeq CF / 0.7 \simeq 0.7$. Lastly, we see that the data in the brightest bin hint at a tentative decrease in $CF$. This is because the most luminous objects in the distribution cannot be interpreted in terms of some even more luminous sources observed under high viewing angles. Therefore, the observed luminosity is close to the intrinsic one, and $CF_{\rm obs}$ tends to $CF$.
We also explored other combinations for the simulation parameters within reasonable values, but in no case we managed to obtain any clear trend.

There are obviously several over-simplifications in this approach, such as the flux limit selection which is much more complex than a bare magnitude cut. Also the assumption of a completely optically thick torus is at odds with the patchy torus models, which admit a non-zero probability of seeing directly through the torus. Lastly, the isotropic and perfectly efficient re-irradiation of the primary continuum by the torus in the infrared is another very simplistic zeroth order assumption. Yet, the lack of a clear decreasing trend, especially in the moderate- to high-luminosity regime, motivates the search for some additional correlation between the $CF$ of the torus and the disc emission.

\section{Methods}
\label{sec:methods}

Here we describe the methods employed to correct the sample photometry for systematic effects and explain the subsequent analyses.

\subsection{The parameter space}
\label{sec:parspace}

\begin{figure*}[h!]
\centering
\includegraphics[width=\linewidth,clip]{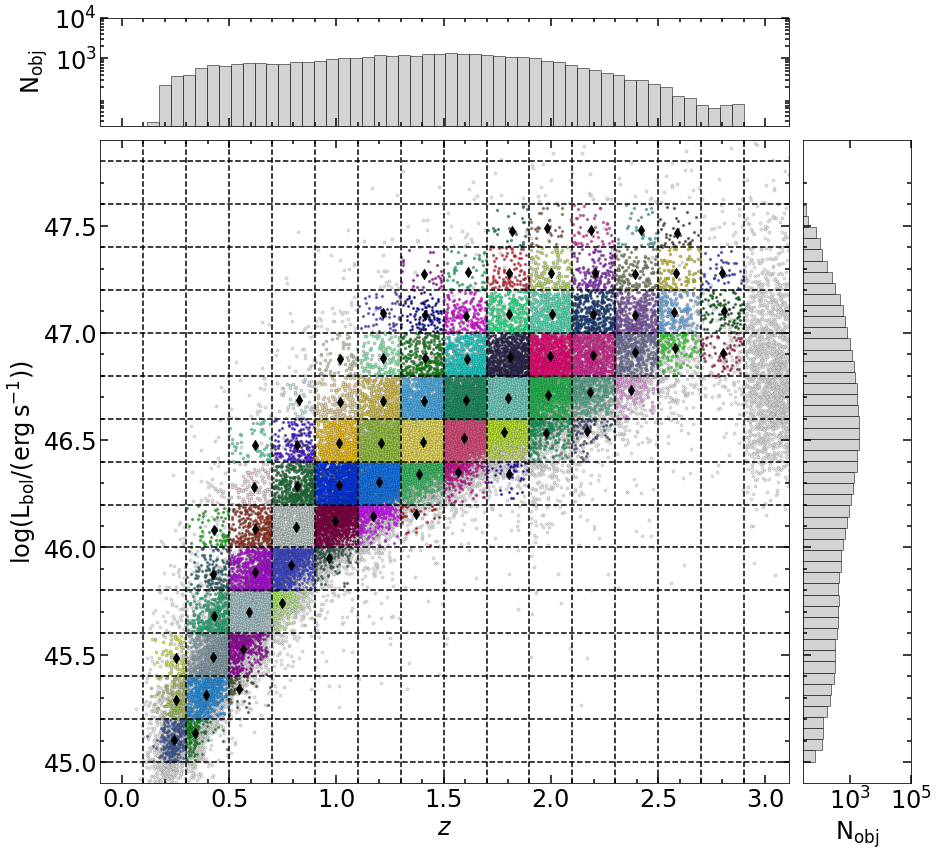}
\caption{$\log L_{\rm bol}$-$z$ plane for the sources surviving the binning selection and the colour cut (see text for details). The final sample spans roughly 2.5 decades in bolometric luminosity and covers up to $z\sim2.9$. Data are randomly coloured if the bin contains more than 30 objects to make the binning clearer. The black diamonds mark the average $z$ and $\log L_{\rm bol}$ values.}
\label{fig:parspace}
\end{figure*}

The key point of this study is the effort to disentangle the effects of the luminosity and the redshift in shaping the NIR emission coming from dusty torus of quasars. To achieve this goal, the sample was analysed in the $\log(L_{\rm bol})$--$z$ plane\footnote{For the sake of a lighter notation, we adopt $\log(L_{\rm bol})$ instead of the more correct $\log(L_{\rm bol}/ (\rm{erg \, s^{-1}})$).}, as shown in Fig. \ref{fig:parspace}, using the reference values for each object tabulated in \citet{wu2022catalog}. We note that when referring to $\log(L_{\rm bol})$ we mean the luminosity evaluated via some bolometric correction in the form $k_{bol} \, \lambda L_{\lambda}$. In the following, we will introduce further quantities designating the integrated accretion disc luminosity.
In order to study the evolution of the torus emission in the parameter space, we divided the distribution in bins of width $\Delta z$=0.20 and $\Delta$log($L_{\rm bol}$)=0.20. The width of the bins is much larger than the typical uncertainty on the redshift, which is of the order of 0.1\%, while it is similar to the typical uncertainty on $L_{\rm bol}$, computed adopting monochromatic bolometric corrections (\citealt{richards2006spectral}).

\subsection{Colour selection}
\label{sec:colour_sel}
\begin{figure}[h!]
\centering
\includegraphics[width=\linewidth,clip]{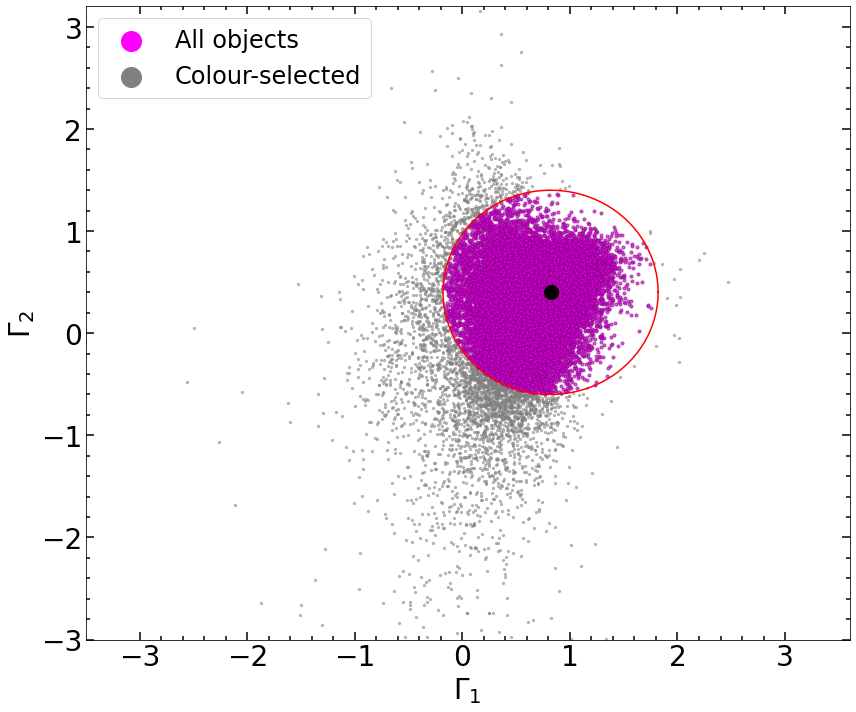}
\caption{$\Gamma_1$--$\Gamma_2$ plane for the sources satisfying the colour cut (magenta dots within the red circle). The black dot marks the average values reported in \citet{richards2006spectral} for blue unobscured quasars.}
\label{fig:colour_cut}
\end{figure}

The importance of the colour cut when aiming to isolate the intrinsic (i.e. not biased by extinction) quasar emission has been discussed in several works (see for instance \citealp{lusso2020}, and references therein). We chose to apply a colour cut to address two main issues: 1) strong contamination from the host galaxy, which is also addressed in Sec. \ref{sec:hgc_corr}, and 2) dust reddening. Both these effects concur to make the observed SED of a quasar flatter than the intrinsic, polluting the broadband information about the relative importance of the torus emission. 
To tackle these issues we adopted an analogous procedure to that described in Section 5.1 of \citet{lusso2020}. In brief, for each object we computed the slope $\Gamma_1$ of a log($\nu$)--log($\nu L_{\nu}$) power law in the rest frame 0.3--1 $\mu m$ range, and the analogous slope $\Gamma_2$ in the
1450--3000 \AA\, range. We selected all the objects in the $\Gamma_1$--$\Gamma_2$ plane residing in the circle with center in the reference values for a typical quasar ($\Gamma_1$=0.82, $\Gamma_2$=0.40; \citealp{richards2006spectral}) and a reddening corresponding to $E(B-V)\lesssim$0.1 assuming an SMC extinction curve (\citealt{prevot84}). The colour selection retained 90\% of the sources, and it is shown in Fig. \ref{fig:colour_cut}. This selection, aimed at avoiding the effects of extinction, does not affect any of our conclusions, as shown in Appendix \ref{app:checks}. The sources surviving this final cut represent the starting sample for our analyses, and amount to 36,000 objects.

\subsection{Photometric corrections}
\label{sec:corrections}

Photometric data convolve the spectral information through band-pass filters. This causes the individual features, easily recognizable in a moderate-quality spectrum, to be instead mixed up in photometric points. Since we are mostly concerned in building the SED of the bare AGN disc and torus emission for each source of the sample, we needed to apply corrections for several effects. In detail, we estimated 1) the contribution from emission lines, 2) the host galaxy emission, 3) the inter-galactic medium (IGM) extinction. We accounted for each of these effects independently. 

Generally speaking, the corrections $\alpha$ applied to each photometric point have the following expression:
\begin{equation}
    L_{f, \, {\rm corr}} = \alpha L_{f}
\end{equation}
where $L_{f} = \int_{f} \lambda L_{\lambda} \big\rvert_{\lambda =\lambda_{\rm obs}} \, S_{\lambda_{\rm obs}} \,d\lambda_{\rm obs}$ is the total luminosity (i.e. $\lambda L_{\lambda}$) integrated through the observed-frame normalised pass-band of the filter $f$. The filter transmission curve $S_{\lambda_{\rm obs}}$ was simulated using the PYPHOT\footnote{\url{https://github.com/mfouesneau/pyphot}} Python package\footnote{The PYPHOT package does not include UKIDSS filters response curves. The $Y$, $J$, $H$, $K$-band responses were simulated adopting the transmission curves available at the Spanish Virtual Observatory (SVO) online facility \url{http://svo2.cab.inta-csic.es/theory/fps3/index.php?mode=browse&gname=UKIRT&gname2=UKIDSS&asttype=}.}.

\subsubsection{Line emission}
\label{sec:line_corr}
The line emission is produced in the gas clumps rapidly moving around the SMBH on pc scales in the BLR in the case of broad permitted lines or on galactic scales in the case of narrow forbidden lines. These features act as contaminants in the photometric data when trying to isolate the AGN continuum emission. In order to subtract the emission line contribution in each filter, we considered the emission of a typical blue quasar, based on the \citet{vandenberk2001} template $T_{\lambda}$. We built a model containing the bare power-law continuum ($T^{c}_{\lambda}$) and another one containing the power-law continuum plus the emission lines, the UV and optical iron pseudo-continua and the Balmer continuum ($T^{c+l}_{\lambda}$). We only corrected for the effect of the emission lines between 1200--7000\,\AA, where the strongest line contribution is expected, while we did not correct for the 9.7\,$\mu m$ feature (expected to be in emission in case of type 1 AGN) in the IR as this falls outside of the wavelength of interest for estimating the $CF$ (see Sec. \ref{sec:lum_proxy}).
For each photometric filter $f$ the emission lines correction $\alpha_{\rm EL}$ was evaluated as

\begin{equation}
    \alpha_{\rm EL}=\frac{\int_{f} \lambda T_{\lambda}^{c}\big\rvert_{\lambda =\lambda_{\rm obs}} S_{\lambda_{\rm obs}} d\lambda_{\rm obs}}{\int_{f} \lambda T_{\lambda}^{c+l}\big\rvert_{\lambda =\lambda_{\rm obs}} S_{\lambda_{\rm obs}} d\lambda_{\rm obs}} \leq 1
\end{equation}

The resulting luminosity in each filter is lower than or equal to the uncorrected one after applying this correction. An example of the result of this correction for a $z=1.0$ quasar is shown in Fig. \ref{fig:corrections}. Here we did not perform a luminosity-dependent emission line correction that would take into account the anti-correlation between the emission line equivalent width and the luminosity (i.e. the Baldwin effect, \citealt{baldwin1977}). Yet, we note that with respect to our results, this is a conservative approach. Indeed if we qualitatively assume weaker emission lines at higher $L_{\rm bol}$ (and $z$), smaller corrections would be implied and, consequently, intrinsically brighter disc luminosities. This, in turn, would translate into even lower torus-to-disc ratios at high $L_{\rm bol}$, thus strengthening our conclusions (see Sec. \ref{sec:lbol_correlation}).

\subsubsection{Host galaxy emission}
\label{sec:hgc_corr}
The observed flux in each filter comes from the combination of the host galaxy and the AGN emission. Luminous AGN ($L_{\rm bol} \gtrsim 10^{46}$ erg s$^{-1}$) are expected to completely outshine their host galaxy. Although the importance of the host-galaxy contamination should have already been minimised by the colour cut (see Sec. \ref{sec:colour_sel}) and the choice of $\log(L_{\rm bol})$=45 as the minimum bolometric luminosity, we aimed at removing any possible residual contamination. We performed this task by subtracting the emission from a typical galaxy, assuming a correction with the form:

\begin{equation}
    \alpha_{\rm HGC} = 1 - \frac{L_{5100\AA,\rm tot} \times f_{\rm HGC} \times \int_{f} \lambda G_{\lambda}\big\rvert_{\lambda =\lambda_{\rm obs}} S_{\lambda_{\rm obs}} d\lambda_{\rm obs}}{L_{f}}   \leq 1
\label{eq:hgc}
\end{equation}

where $L_{5100\AA, \rm tot}$ is the total luminosity at rest-frame 5100\,\AA\, evaluated by interpolating the available photometry, $f_{\rm HGC}$ is the host-galaxy fraction evaluated at 5100\,\AA\,, $G_{\lambda}$ is the spectrum of a typical galaxy and $L_{f}$ is the photometric luminosity through the filter $f$.
As for the galaxy spectrum we chose an elliptical galaxy template from \citet{assef2010low}, scaled by its 5100\,\AA\, luminosity. The host-galaxy fraction is a function of the AGN luminosity: to account for this we took advantage of the prescription described in \citet{shen2011}, which parameterises $f_{\rm HGC}$, here adapted as:
\begin{equation}
    \frac{L_{5100, \rm host}}{L_{5100, \rm QSO}} = 0.8052 - 1.5502 x + 0.9121 x^{2} - 0.1577 x^{3}
\end{equation}
where $x$=$\log(L_{5100\AA, \rm tot}/(\rm erg \, s^{-1}))-44$.
No correction was applied for objects with $x>1.053$, that represent 80\% of the sample. Although \citet{shen2011} evaluated their correction using only $x\geq0$, we extrapolated their parameterisation also below this value. We note that the fraction of quasars below this value is negligible ($\sim 0.03\%$) and these objects are mostly discarded as residing in low counts bins in the subsequent analysis. We therefore do not expect this effect to seriously hamper our results. In a few cases ($\sim$1\% of the objects suitable for the host-galaxy correction) the optical photometric data presented a steep decline towards redder wavelengths, incompatible with the adopted galaxy template. In these cases the AGN emission would be outshined by the host, giving $\alpha_{\rm HGC}<0$. Whenever this condition occurred, no correction was applied. Although, for the sake of completeness, we included this correction, we note that this was applied only to $\sim$20\% of the sources in the sample, which reside in the low-$L_{\rm bol}$ low-$z$ corner of the parameter space, and will not be included in the following analyses due to the lack of faint counterparts at high $z$.

\subsubsection{IGM extinction}
\label{sec:igm_corr}
At wavelengths shorter than the Lyman $\alpha$ the rest frame emission of each quasar is attenuated by the intergalactic \ion{H}{i} absorption, creating both the so-called Lyman $\alpha$ forest in line absorption and a drop in the continuum below the Lyman limit at $\lambda < 912$\,\AA\, (e.g. \citealt{moller1990lyman}). Because of the IGM absorption, the observed luminosity $\lambda L_{\lambda, \rm obs}$ is lower than the intrinsic $\lambda L_{\lambda, \rm int}$ by a wavelength- and redshift-dependent transmission factor $T(\lambda, z)$, so that $\lambda L_{\lambda, \rm obs} = T(\lambda, z) \times \lambda L_{\lambda, \rm int}$. The transmission is linked to the effective \ion{H}{i} Lyman series and Lyman continuum photons optical depth $\tau(\lambda, z)$ by $T_{\lambda}(z) = e^{-\tau(\lambda, z)}$. The goal of this correction is to increase the flux in the photometric bands affected by absorption by the IGM ($\alpha_{\rm IGM}\geq1$), in order to recover the intrinsic ultraviolet (UV) and extreme ultraviolet (EUV) flux. The IGM absorption correction $\alpha_{\rm IGM}$ can thus be evaluated by assuming the knowledge of the shape of the EUV continuum as:

\begin{equation}
    \alpha_{\rm IGM}=\frac{\int_{f} \lambda L_{\lambda}\big\rvert_{\lambda=\lambda_{\rm obs}} \, S_{\lambda_{\rm obs}} d\lambda_{\rm obs}}{{\int_{f} \lambda L_{\lambda} T_{\lambda}(z)\big\rvert_{\lambda=\lambda_{\rm obs}} \, S_{\lambda_{\rm obs}} d\lambda_{\rm obs}}} \geq 1
\end{equation}

\begin{figure*}[h!]
\begin{subfigure}{0.5\textwidth}
\includegraphics[width=\linewidth,clip]{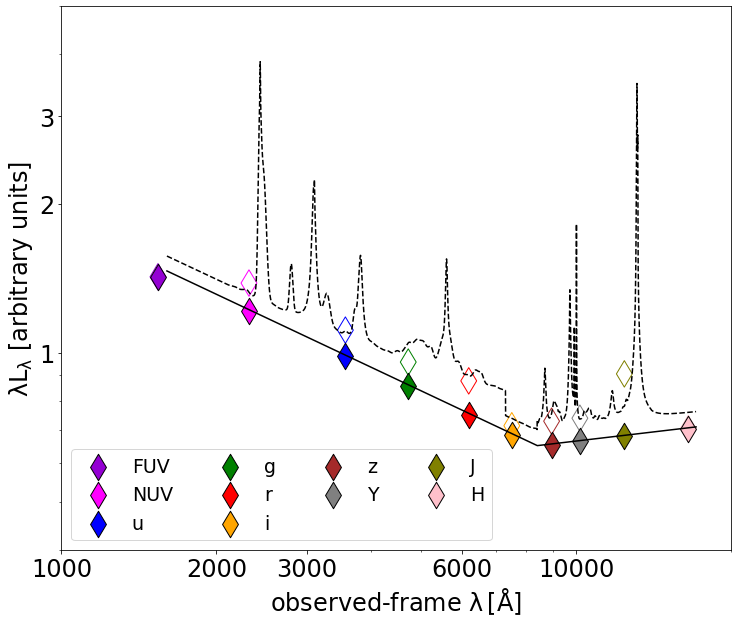}
\label{fig:corrections_1}
\end{subfigure}
\hspace{0.02\textwidth}
\begin{subfigure}{0.5\textwidth}
\includegraphics[width=0.95\linewidth]{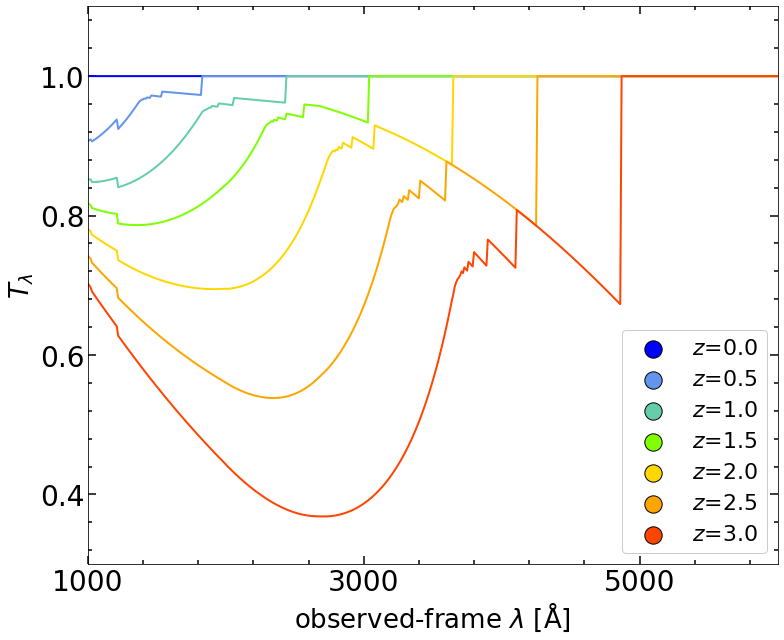}
\label{fig:corrections_2}
\end{subfigure}
\caption{\textit{Left:} pass-band integrated luminosity of the template with (empty diamonds) and without (filled diamonds) emission lines for a $z=1.0$ quasar, according to the \citet{vandenberk2001} template. \textit{Right:} IGM transmission curves at different redshifts colour-coded as described in the legend.}
    
\label{fig:corrections}
\end{figure*}

Here we assumed the intrinsic EUV continuum to follow the behaviour described in \citet{lusso2015}, i.e. $L_{\lambda}\sim \lambda^{\alpha_{\lambda}}$ with $\alpha_{\lambda} = -0.3$. We estimated the effect of the IGM absorption at different redshifts by taking advantage of the calculations described in \cite{inoue2014updated}. There, the absorption is considered to be delivered mainly by two components, the Ly$\alpha$ forest component and the damped Ly$\alpha$ systems, acting on different column density regimes. In order to be conservative and apply the minimal correction, we only included the Ly$\alpha$ forest in our corrections. Examples of the transmission curves are reported in the right panel of Fig. \ref{fig:corrections}.

\begin{figure*}[h!]
\begin{subfigure}{0.5\textwidth}
\includegraphics[width=0.95\linewidth]{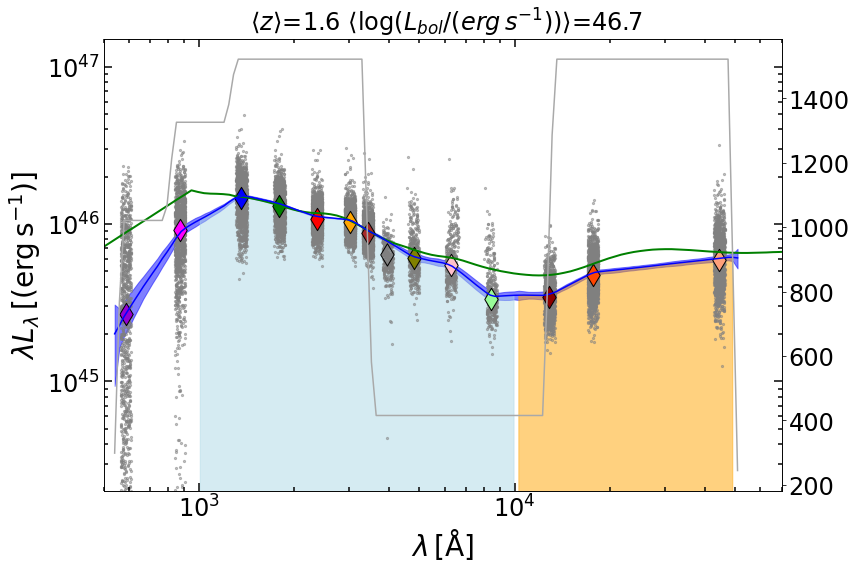}
\end{subfigure}
\hspace{0.02\textwidth}
\begin{subfigure}{0.5\textwidth}
\includegraphics[width=0.95\linewidth]{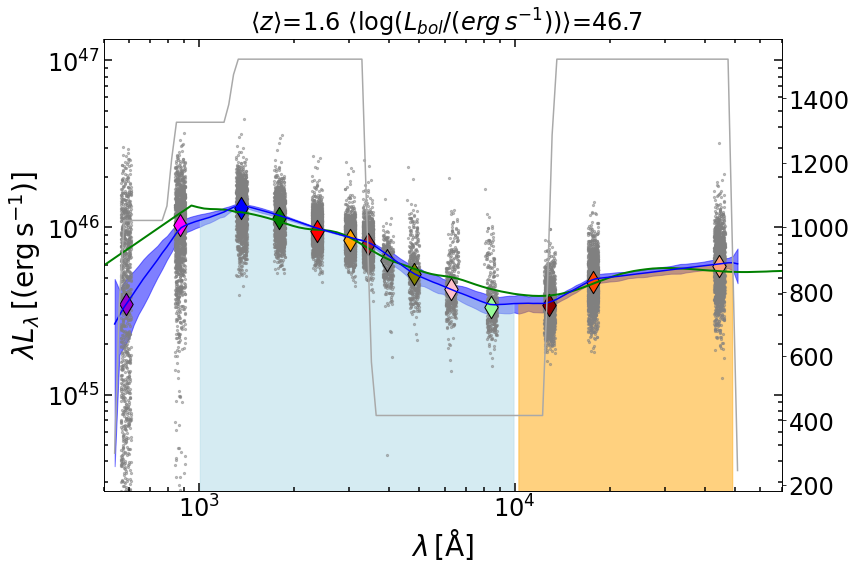}
\end{subfigure}
\caption{Example of average SEDs produced employing the uncorrected (left) and corrected (right) photometry (blue). The effect of the corrections applied is to increase the flux bluewards of the Ly$\alpha$, mitigate the dip around 1 $\mu$m and reduce the optical and UV \ion{Fe}{ii} bumps. Also the strong H$\alpha$ emission in the observed-frame $H$ band (pink) is reduced in the corrected data-set. The grey clouds of points are made of the photometric data used to produce the SED, the coloured diamonds (same colour-code as in left panel of Fig. \ref{fig:corrections}) are the mean luminosities. The solid grey line represents the number of objects contributing to each spectral channel. The average quasar SED from \citet{krawczyk2013mean}, scaled to the 3000\,\AA\, total luminosity, is shown in green as a comparison. The azure and orange shaded areas represent the regions where the $L_{\rm opt}$ and the $L_{\rm NIR}$ terms in the covering factor proxy $R$ are respectively evaluated (see Sec.\ref{sec:lum_proxy}). The average $z$ and $\log(L_{\rm bol})$ of the parameter space represented by the SED are denoted in the top string.}
    
\label{fig:sed_ex}
\end{figure*}

\subsection{Building the average SEDs}
\label{sec:make_seds}

Here we describe the technique adopted to build the average SEDs employed in this work. The procedure is the same for building both the average SED in each bin of the $L_{\rm bol}-z$ parameter space  
and those obtained along rows or columns of the parameter space.
After performing the mentioned corrections and quality cuts, we proceeded to gather all the photometric information in the selected region of the parameter space. For each source the photometric pairs ($\log\lambda_{\rm rest}$, $\log\lambda L_{\lambda}$) were rebinned onto a common wavelength grid made of 300 logarithmically equispaced points between $10^{1.5}-10^{6}$\AA\,, building a \textit{photo-array} for each object.

Not all the objects have photometric coverage between the $Y$ and the $K$ band. A direct interpolation between the SDSS-$z$ and the WISE-$W1$ bands could overestimate the actual SED which generally shows a dip around 1$\mu m$ (see e.g. \citealp{richards2006spectral, elvis2012spectral, krawczyk2013mean}). Because of this, in each composite SED the flux between the last SDSS filter and W1 was computed using only sources with actual UKIDSS and/or 2MASS data, rather than by interpolation.

The average SED and the relative uncertainty in each bin were then produced by adopting a bootstrap resampling. In brief, a number of \textit{photo-arrays} equal to the total number of objects in the bin is randomly drawn. The \textit{photo-arrays} were scaled by the monochromatic 3000 \AA\, luminosity and an average stack is produced by considering the median luminosity in each spectral channel. This scaling wavelength was chosen for being covered by SDSS photometry in most cases. This procedure is repeated with new random draws until the desired number of stacks (N=100) is reached. The final composite SED is made by the median of the 3$\sigma$-clipped distribution of fluxes in each spectral channel, while the uncertainty is set by the standard deviation. The final SED is scaled so that the average luminosity and the relative uncertainty at 3000 \AA~($\lambda L_{\lambda}\big\rvert_{\lambda = 3000\rm{\AA}}$) are respectively equal to the mean and the standard error on the mean $L_{3000\AA}$ in the bin. An example of the result of this procedure, also including the effect of the corrections to the photometry, is shown in Fig. \ref{fig:sed_ex}.

\subsection{Defining a proxy for the covering factor}
\label{sec:lum_proxy}

The emission of the dusty torus in AGN is expected to peak between $\sim$3--30 $\mu m$, with noticeable differences depending on the covering factor ($CF$), the optical depth ($\tau$) and the inclination of the line of sight (see e.g. \citealp{nenkova2008agn, shi2014infrared, garcia2017mid}). 
In order to precisely define the covering factor, a complete description of the SED, representing the whole accretion disc and torus emission from the extreme UV to the far infrared, is needed. Unfortunately, this is not generally feasible, especially for large samples like ours. 

Several past works have adopted different infrared and UV/optical proxies, respectively, for the torus and AGN luminosity terms in the $CF$ expression. Traditionally, the ratio $R$ between the monochromatic luminosities at some infrared and UV/optical wavelengths (e.g. \citealt{maiolino2007dust, treister2008measuring}) or integrated over the available wave-bands (e.g. \citealt{cao2005dust, gu2013evolution, ralowski2023covering}) has been widely employed. 

A more refined approach to the problem of evaluating the covering factor requires taking into account other factors such as the optical thickness of the torus and its shape. For instance, an optically thick torus is expected to have a lower $CF$ (see e.g. Section 6 in \citealt{lusso2013obscured}). Moreover, it has been shown by \citet{stalevski2016dust} that the clumpiness of the dusty torus in combination with the anisotropic emission of the accretion disc leads to the underestimate (overestimate) of low (high) $CF$ in type 1 AGN. To correct for these effects, they propose a polynomial expansion of $CF$ in terms of $R$. 

The wavelengths covered by our data do not consent to properly sample the mid- and far-infrared tail of the torus emission. Similarly, the low coverage in the far-UV combined with the effect of possible residual (and redshift-dependent) IGM absorption, not completely corrected by our procedure described in Sec. \ref{sec:igm_corr}, does not guarantee a complete description of the BBB. Because of this, we rely on observational proxies for the $CF$ based on different combinations of UV/optical and near-infrared (NIR) luminosities. Throughout this work, we adopted the ratio $R=L_{\rm NIR}/L_{\rm opt}$\footnote{Here we highlight the difference between $L_{\rm bol}$ and $L_{\rm opt}$. The former is evaluated via a bolometric correction, and is only used to place every source in the parameter space. The latter is instead computed by directly integrating the SED between 1,000 \AA\, and 1 $\mu$m and is adopted to estimate the proxies of $CF$.} as a proxy for the intrinsic $CF$, with $L_{\rm NIR}$ being the integrated luminosity of the average SEDs between 1--5 $\mu$m and $L_{\rm opt}$ that between 1,000 \AA\, and 1 $\mu$m.

In the NIR, the WISE coverage at the redshifts sampled in this work only allows us to access the hot dust emitting in the range between roughly 1--5 $\mu$m. This spectral region can be affected by stellar emission within the host galaxy (e.g. \citealt{podigachoski2016starbursts}), the infrared tail of the accretion disc and/or sources with strong contributions by hot graphite dust (e.g. \citealt{mor2009dusty}). In our case, 
the selection of luminous AGN (the sample mean is $\log(L_{\rm bol})$ is 46.4 erg s$^{-1}$), as well as the correction for the host galaxy emission, help in mitigating the effect of the possible stellar light contribution. In addition, we visually inspected all the average SEDs to check that the 1 $\mu$m emission was dominated by the disc, rather than by the peak of the peak of the stellar bulge in the host, thus minimising the possible mixing of the components. On the other hand, \citet{stalevski2016dust} showed that, within their model, the 1--5 $\mu$m band has the favourable aspect of having almost no dependence on optical depth, and on the other parameters explored therein, such as the clumpiness of the torus and the inclination angle.

For what concerns the AGN term in the $CF$ definition we employed $L_{\rm opt}$, defined as the integrated UV/optical BBB emission from 1,000 \AA\, emission up to 1 $\mu m$ in the average SEDs, including only the bins where the peak of the disc emission was clearly observed in the data. If the UV/optical SEDs peaked at 
significantly different wavelengths at different redshfits and luminosities this choice would provide unreliable results. We here 
rely on the evidence that the SED of quasars in different ranges of luminosity and redshift appear to universally peak around 1,000 \AA\,  (see e.g. \citealt{zheng1997composite, telfer2002rest, shull2012hst, stevans2014hst,  cai2023universal}). In Appendix \ref{app:checks} we also tested how a variable lower integration limit, following the peak of the disc emission, could affect our results, but found no significant differences. Furthermore, we present the results obtained also adopting other proxies for $CF$, obtained with different combinations of $L_{\rm NIR}$ and $L_{\rm opt}$ as well as with alternative datasets. In all the cases, the anti-correlation between $R$ and $L_{\rm bol}$ and the non correlation between $R$ and $z$ are recovered.

The uncertainty on $R$ was evaluated by producing 100 mock SEDs for each object, by randomizing the flux in each spectral channel adding a Gaussian deviate using the uncertainty on the final SEDs.

\subsection{Reaching the rest-frame 5 $\mu$m}

In this work, we limited our analysis to $z<3$ so that our sample reaches a high degree of completeness at $i<19.0$. This implies that the rest-frame 5 $\mu$m falls within the W3 filter up to $z\sim2.3$. Above this value, an increasing portion of the rest-frame 1--5\,$\mu$m emission is sampled in the observed-frame by the W4 filter, which has been shown to be prone to overestimating the intrinsic flux when compared with instruments with similar passbands (e.g. Spitzer MIPS24,  \citealt{ralowski2023covering}). We confirmed this issue by cross-matching the W4 photometry with that of Spitzer MIPS 24 $\mu m$ for a sample of common sources. In agreement with the previous findings we confirm that W4 systematically overestimates the flux in the low SNR regime. A more quantitative discussion on this issue is presented in Appendix \ref{app:checks}. Nonetheless, at $z<3$ the emission between 1--5 $\mu m$ is mostly covered by W1, W2, W3. In the cases where the rest-frame 5-$\mu m$ emission falls redwards of the W3 filter we extrapolated the SED using the W2 and W3 photometric points using a power law. We estimated how much this prescription could underestimate the actual value of $CF$ by performing the following exercise. We considered the average SEDs where 5 $\mu m$ falls redwards of the central wavelength of W3 (12.3 $\mu$m). We assumed a typical NIR quasar spectrum represented by the template built in \citet{hernan2016near}, and matched it to the average SED at the rest-frame W3 wavelength, using it to cover the rest 5-$\mu m$ emission. We then evaluated $R$ from the template-matched and from the extrapolated SED. The typical difference is at the per cent level. We also note that this is a conservative approach, because of the actual width of the W3 filter. For instance, at 16.3 $\mu$m the filter response is still about $\sim$50\%.

\section{Results}
\label{sec:results}

In this Section we report the results of the analysis regarding the evolution of NIR SED and $R$ with $z$, $L_{\rm bol}$ and $\lambda_{\rm Edd}$. Additionally, we compare the average SED built upon the whole sample with others in literature, and present how the information about the dusty torus gathered in this work can give constrains on the mechanism responsible for obscuration in the X-rays.

\subsection{Evolution of the average SED with $z$ and $L_{\rm bol}$}
\label{sec:nir_evolution}
A key question when trying to understand the observed properties of the obscuring torus in AGN is whether and how they change as a function of the AGN parameters ($L_{\rm bol}$, $M_{\rm BH}$) and of the cosmic epoch.
For what concerns the luminosity evolution, the sub-linear relations claimed by different authors (e.g. \citealt{maiolino2007dust, ma2013covering, gu2013evolution}) between the optical and infrared luminosities evaluated at several pairs of wavelengths, imply a relative decrease of the NIR luminosity for increasing optical and UV luminosity. This feature, as already mentioned, fits in the physical picture where the torus recedes for increasing accretion disc luminosity.
From a cosmological perspective, the IR SED of high-redshift, non dust-deficient quasars (\citealt{leipski2014spectral}) has been shown to closely resemble, on average, that of more local samples (e.g. \citealt{elvis1994atlas, richards2006spectral}), thus hinting at the fact that a non-evolution of the IR SED is actually possible. Here we aim to explore systematic changes in the shape of the NIR SED as a function of both $L_{\rm bol}$ and $z$ separately.

\begin{figure*}[h!]
\begin{subfigure}{0.5\textwidth}
\includegraphics[width=0.95\linewidth]{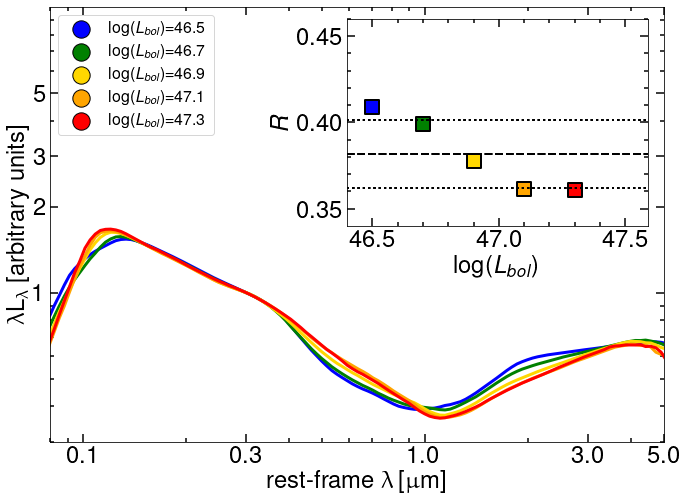}
\end{subfigure}
\hspace{0.02\textwidth}
\begin{subfigure}{0.5\textwidth}
\includegraphics[width=0.95\linewidth]{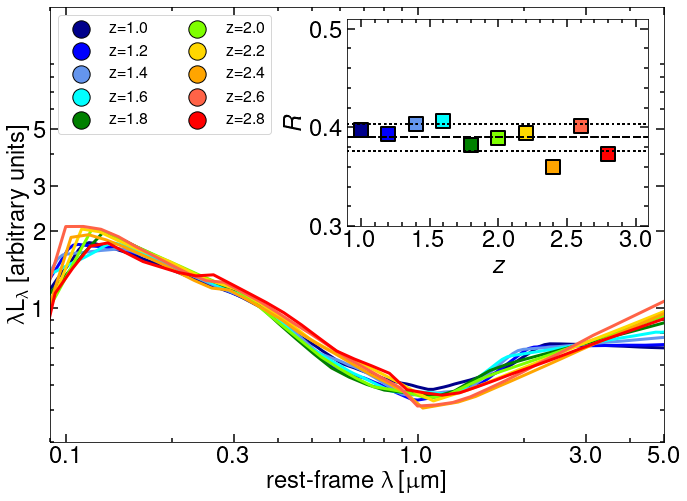}
\end{subfigure}
\caption{\textit{Left:} Sequence of average SEDs for increasing luminosity, scaled by their total luminosity between 1,000\,\AA\, and 1 $\mu$m. The trend of decreasing NIR luminosity for increasing optical luminosity is apparent, although the small dynamical range, and it is testified by the decreasing $R$ shown in the inset. Here the mean value (standard deviation) is shown as a black solid (dashed) line. \textit{Right:} Sequence of average SEDs for increasing redshift, scaled by their total luminosity between 1,000\,\AA\, and 1 $\mu$m. The different sampling of the rest-frame emission leads to slightly different SEDs. However, on average $R$ does not show any evolutionary trend, as we do not observe systematic departures from the mean value.}
    
\label{fig:Lz_seq}
\end{figure*}

For this purpose, we performed the following exercise: we considered all sources residing in the region of the parameter space comprised between $46.4 \leq \log(L_{\rm bol}) \leq 47.4$ and $1.0 \leq z \leq 2.8$ ($\sim$21,000/36,000 quasars) and proceeded to produce composite SEDs for each luminosity or redshift bin while integrating along the other axis, following the procedure described in Sec. \ref{sec:make_seds}. We chose this subspace because, within this region, the coverage of the parameter space is fairly homogeneous, although the luminosity tends to slightly increase with redshift. For example, the average $\log(L_{\rm bol})$ at $z\sim$1.0 is 46.58, while it reaches 47.06 in the highest redshift bin at $z\sim2.8$. However, including also sources below $z$=1.0, the mismatch in terms of luminosity between the low- and the high-redshift composites, would be even worse, becoming as high as $\sim$1.5 dex.
We then employed these average SEDs (shown in Fig. \ref{fig:Lz_seq}) to test the evolution of the NIR SED shape with $z$ and $L_{\rm bol}$ separately. We also highlight in the inset panels of Fig. \ref{fig:Lz_seq} the values of $R$ derived from these composite SEDs. It is easy to assess the receding NIR SED with increasing luminosity, a feature also testified by the decreasing $R$ shown in the inset plot, albeit the limited dynamical range of only one order of magnitude. The composite SEDs in $z$ bins exhibit slightly more pronounced SED to SED variations, particularly at $\sim$2 $\mu$m where a turnover appears, mostly in the data at $z\lesssim 1.6$. Yet, we note that producing average SEDs at different $z$ introduces the technical issue given by the uneven sampling of the rest-frame spectrum, which, in the NIR, is exacerbated by two factors: the first is the scarce sampling due to the presence of only the three WISE filters W1, W2, W3, and the second being the non-power law shape of the continuum. The combination of these factors likely leads to the seeming differences in the NIR of the average SEDs at different $z$. However, such effect is not present in the SEDs at fixed luminosity where, instead, the wide range of redshifts allows for a finer sampling of the rest-frame spectra.

\begin{figure}[h!]
\centering
\includegraphics[width=\linewidth,clip]{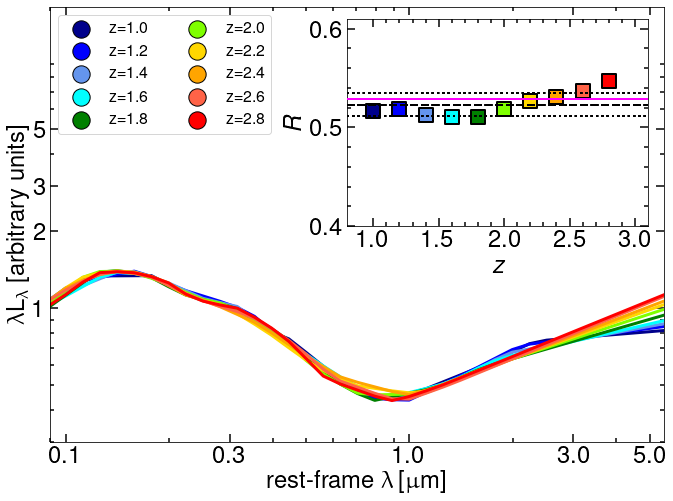}
\caption{The composite SEDs from the simulated photometry of the template in the same redshift bins of the actual data. Here the magenta line marks the actual $R$ value estimated using the total spectral template, rather than the photometry.}
\label{fig:sim_zseq}
\end{figure}

With the end of verifying whether the SEDs in bins of fixed $z$ are really consistent with an intrinsically constant NIR SED not varying with the redshift, we performed an additional test: we created an intrinsic FUV-to-NIR SED by joining three spectral templates, namely the \citet{zheng1997composite} template for the FUV, the \citet{selsing2016x} for the optical and the \citet{hernan2016near} for the NIR. The choice of the \citet{selsing2016x} template, rather than the \citet{vandenberk2001}, was mainly driven by the fact that the former is based on luminous quasars, hence the optical band is almost unaffected by the host galaxy, as it is likely for the objects in the subsample considered here. In this case we opted for a NIR spectral template, rather than a photometric SED, in order to include the effect of the emission lines at different $z$, as we did not subtract NIR emission line features from our photometric data set. As done for each of the sources in the parent sample, we removed the contribution from prominent broad emission lines, as well as the UV and optical \ion{Fe}{ii} pseudo-continua. We then shifted the spectral template to the same redshifts where the average SEDs had been produced, and convolved it with the filters of the surveys employed in this work. Lastly, we evaluated $R$ in the same way as it was estimated from the actual SEDs. In the cases where the observed-frame photometry did not cover the rest-frame 5 $\mu$m, we resorted to extrapolation. The result of this further test is shown in Fig. \ref{fig:sim_zseq}. Here we highlight two noticeable features: the first is that the turnover at $\sim$2 $\mu$m, present in the composite SEDs below $z\sim$1.6, is also observed in the simulated SEDs at the same redshift, where both the rising and the flat regions of the NIR bump (consistent with a blackbody emission at $T\sim 1200$ K, \citealt{hernan2016near}) are sampled by W1, W2 and W3. As the redshift increases, only the rising part of this bump is sampled by the WISE filters, hence the power-law extrapolation. This effect is shown in detail in Appendix \ref{app:nir_bump}, where we demonstrate how the interpolation (and extrapolation) across the filters of the template varies with the redshift. Secondly, we note that, although $R$ is slightly different in absolute value with respect to that observed in the actual data\footnote{This difference might be due to the fact that the one or more of the templates joined to build the total composite spectrum are not fully representative of the subsample under analysis. Nonetheless, the main concern here is not about the absolute value of $R$, but rather how the wave-bands shifts affect its estimates.}, it does not show any conspicuous trend, notwithstanding the different sampling wave-bands. The dispersion in $R$ is close to consistency with a single NIR SED at all redshifts: the standard deviation of $R$ estimated in the simulated SEDs (0.010) is indeed remarkably similar to that of actual SEDs (0.012), thus leaving little room for further effects causing the observed dispersion. As a comparison term, the standard deviation of $R$ at different $L_{\rm bol}$ reaches 0.021 in just one order of magnitude, while also showing an obvious decreasing trend (see the inset plot of the left panel in Fig.~\ref{fig:sim_zseq}).

To sum up, the average SEDs produced at fixed $L_{\rm bol}$ confirm the trend of a weaker NIR emission for increasing optical luminosity. This is clearly observed even employing a relatively small dynamical range (1 dex). We also argue that, although some SED-to-SED variations, the average SEDs at different $z$ are consistent with an intrinsically similar rest-frame NIR SED sampled at different observed-frame wave-bands, rather than with an actual evolution with $z$. 
There are obviously other effects, here not taken into account, which might contribute to the SED-to-SED differences. For instance, in the real data, the average luminosity increases slightly in each redshift bin, as already noted, therefore influencing the NIR emission accordingly. Furthermore, the differences in luminosity are likely accompanied by minor changes in the emission line strength, which are not taken into account by our fixed NIR template.

\subsection{Analysis of the correlation between $CF$, $L_{\rm bol}$ and $z$}
\label{sec:lbol_correlation}
The evolution of the NIR SEDs with $L_{\rm bol}$ can be epitomised in the (sub-linear) anti-correlation between the $CF$ and the AGN luminosity. The fraction of the solid angle covered by the obscuring torus diminishes for increasing luminosity. Here we aim at quantifying the respective effects of the luminosity and redshift in driving the evolution of $CF$.

The main axes of our parameter space, i.e. the bolometric luminosity and the redshift are correlated for both physical and observational reasons. Indeed, more luminous AGN, or quasars, are generally observed at redshifts corresponding to the so-called ``quasar epoch'' at $z\sim 2-3$, while, observationally the SDSS is a flux limited survey. We therefore aimed at employing statistical tools to assess the degree of correlation between our proxies of the $CF$ and the main parameters that could account for this effect. In particular, we adopted a methodology similar to that used in \cite{croom2002correlation}, who studied the independent effects of luminosity and redshift on the evolution of the equivalent width in the quasar population. In particular they used both a partial correlation analysis (PCA, which we will introduce shortly), and performed linear fits between the parameter of interest and the equivalent width, while keeping the other fixed.

\begin{figure}[h!]
\centering
\includegraphics[width=\linewidth,clip]{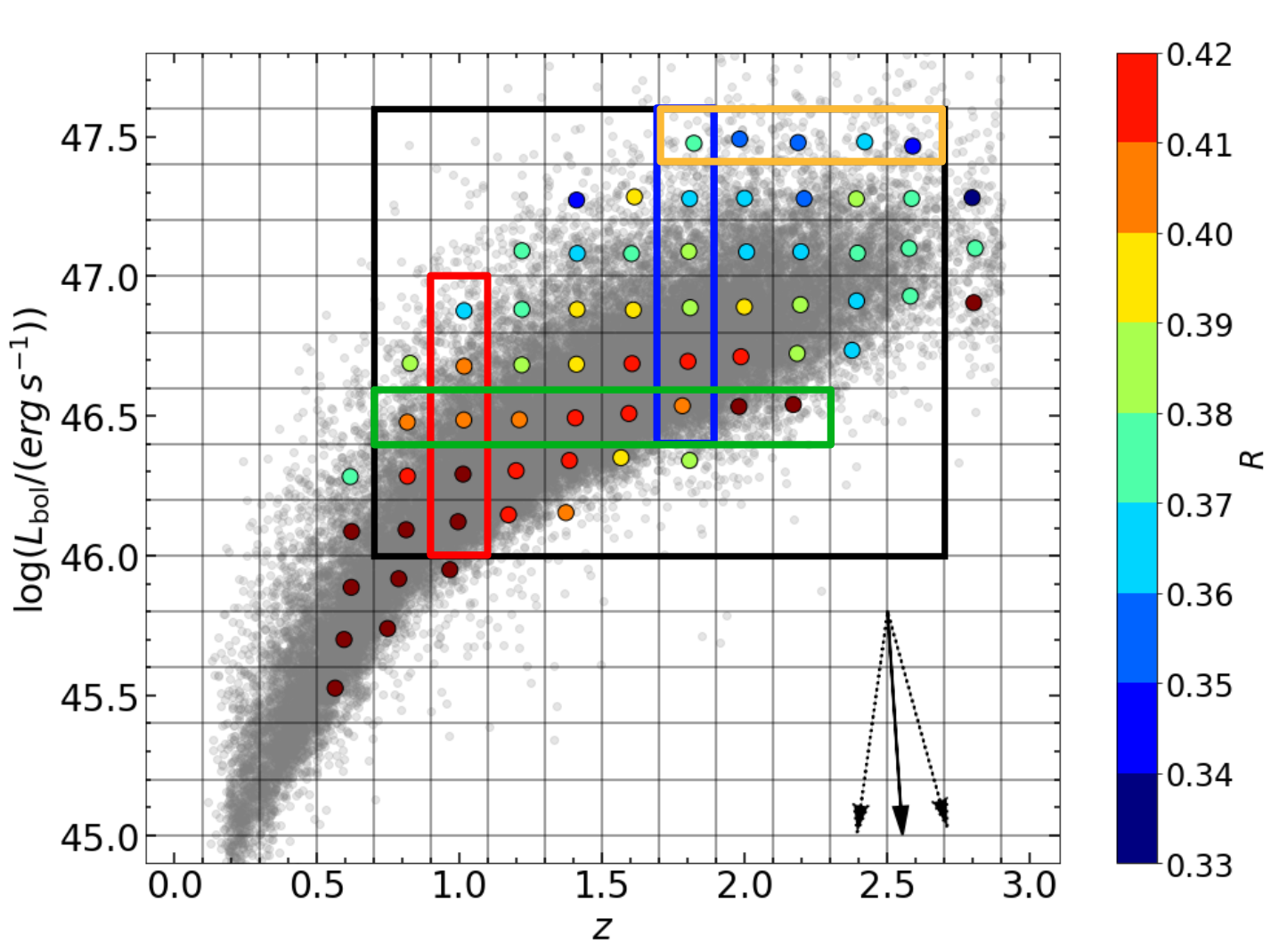}
\caption{$\log(L_{\rm bol})$--$z$ plane with $R$ in colour-code. The black arrow points in the direction given by the PCA, evaluated in the bins within the black rectangle, while the dashed arrows show the uncertainty on the direction. The coloured rectangles highlight the bins of the parameter space shown in Fig. \ref{fig:ex_correlations}.}
\label{fig:parspace_cf}
\end{figure}

The first approach in this analysis was to estimate the degree of correlation between $R$ and the main parameters via a PCA. This tool allows us to disjointly evaluate the degree of correlation between $R$ and one of the main parameters, while keeping the other fixed (see e.g. \citealt{bluck2022quenching, baker2023molecular}). This kind of analysis requires a monotonic dependence of $R$ on both the main parameters. This is the case for the $\log(L_{\rm bol})-R$ relation, while the correlation between $R$ and $z$ appears weaker, as also testified by the correlation indices explored below. In particular, given the three or more quantities (A, B, C) we wish to test the correlation on, the partial correlation coefficient between A and B while keeping C fixed can be expressed as:

\begin{equation}
    \rho_{\rm AB \mid C} = \frac{\rho_{\rm AB} - \rho_{\rm AC} \, \rho_{\rm BC}}{\sqrt{1 - \rho_{\rm AC}^2} \, \sqrt{1 - \rho_{\rm BC}^2}},
\label{eq:pca}
\end{equation}
where $\rho_{\rm XY}$ denotes the Spearman rank correlation index between quantity X and Y. We used the PCA values to define a gradient arrow in Fig. \ref{fig:parspace_cf}, whose inclination is defined as:
\begin{equation}
    \tan \theta = \frac{\rho_{R \, L_{\rm bol} \mid z}} { \rho_{R \, z \mid L_{\rm bol}}}
\end{equation}

The arrow shows the direction of strongest variation in the tested parameter. The more aligned the arrow with an axis, the stronger the correlation with that quantity. The PCA arrow is almost completely ($\theta = -87^{\circ}$) inclined towards lower $\log L_{\rm bol}$ values, signifying that the increasing bolometric luminosity is ultimately responsible for the the decrease of the covering factor. We performed this analysis in the region marked by a thick black rectangle in Fig. \ref{fig:parspace_cf}, which is made of 56 bins with more than 30 sources per bin. This choice avoids the low-luminosity tail at redshift below $z=$0.7, where $R$ reaches its maximum values. We note, however, that the inclusion of these points would not alter any of the results presented here.
We also evaluated the uncertainty on this parameter via a bootstrap resampling of the values of $R$. We computed $\theta$ 1,000 times, each time using a random subsample made of only 30/56 bins in the PCA region, allowing for re-immission. The uncertainty on the PCA inclination was evaluated as the standard deviation of the distribution of $\mid \theta \mid$.
For the sake of completeness we also report in Table \ref{tbl:tbl_corr_1} the partial correlation coefficients and the respective p-values associated to the null hypothesis that there is not any correlation between the variables. In Fig. \ref{fig:parspace_cf} we show $R$ derived from the SED in each bin of our parameter space in colour-code superimposed to the whole sample. We also plot the arrow representing the main direction of the PCA, which clearly indicates $L_{\rm bol}$ as the main driver of the evolution of $R$.

As a further test, we also employed a similar procedure based on the use the Kendall $\tau$ index of correlation (\citealt{kendall1938new}) instead of the Spearman one, as it provides more robust results in case of outliers. The partial correlation evaluated using the Kendall $\tau$ has the same form highlighted in Eq. \ref{eq:pca} with $\rho$ being substituted by $\tau$ (\citealt{akritas1996test}). In this case, we adopted a permutation test to compute the associated p-value (see e.g. \citealt{tibshirani1993introduction}). In brief, given three parameters, we wish to test the correlation on (A, B, C). We shuffled the array containing the parameter A, to simulate a non correlation (random association) between both A and B and A and C, while keeping the relation between B and C unaltered. Then, we proceeded to evaluate the partial correlation coefficient $\tau_{AB\mid C}$ for this simulated dataset. This procedure was repeated 10,000 times, producing the test distribution, and the p-value was estimated as the two-tailed fraction of simulated datasets where $\tau_{AB\mid C}$ exceeded the observed value. The resulting values are reported in Table \ref{tbl:tbl_corr_1}.

Additionally, as another independent check on the correlation strength we performed the regression of linear forms between $R$ and one of the two parameters, while keeping the other fixed. Again, this was performed in the rows or columns of the parameter space within the same region where the PCA was evaluated. Also in this case, the inclusion of the high $R$ bins at $\log(L_{\rm bol})<$ 46.0 would make the $R$-$\log(L_{\rm bol})$ anti-correlation even stronger.
In more detail, we adopted functional forms as:

\begin{align}
   & i) \,\, R = a_0 + a_1 \, (z - z_0) \\
   & ii) \,\, R = a_0 + a_1 \, [\log(L_{\rm bol}) - L_0],
\label{eq:eqtns}
\end{align}

respectively for the $R-z$ and the $R-\log(L_{\rm bol})$ relations, with $z_0$ and $L_0$ being the mean of the $z$ or $\log (L_{\rm bol})$ values upon which the regression was performed. In this case the slope of the line gives information about the strength of the correlation between $R$ and one of the two parameters while keeping the other fixed. 
We report the average slopes and offsets of the best fit relations in Table \ref{tbl:tbl_corr_2}.
We remark that both the relations are merely empirical efforts to describe the correlation between the parameters of our space and the proxy that we adopted for the actual covering factor. These results are used as statistical tools of comparison, while a physical description of the evolution of the covering factor with any of these parameters would require a detailed modelling, beyond the scope of this work.

\begin{figure*}[h!]
\begin{subfigure}{0.5\textwidth}
\includegraphics[width=0.95\linewidth]{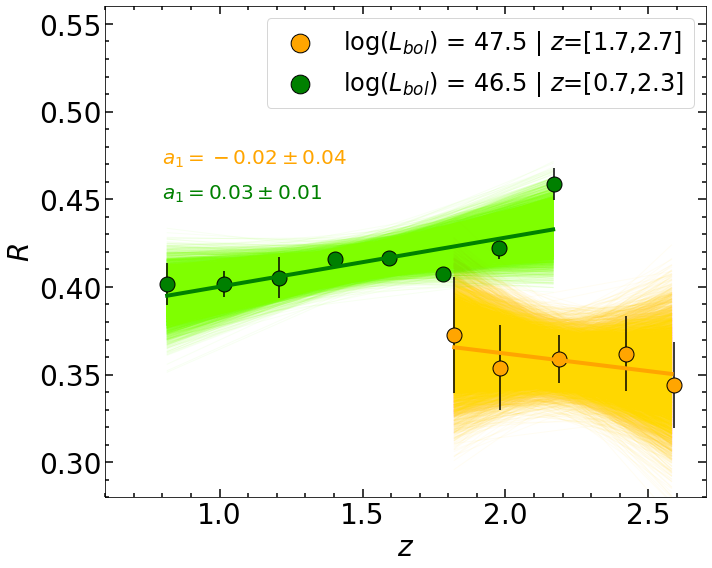}
\label{fig:z_cf}
\end{subfigure}
\hspace{0.02\textwidth}
\begin{subfigure}{0.5\textwidth}
\includegraphics[width=0.95\linewidth]{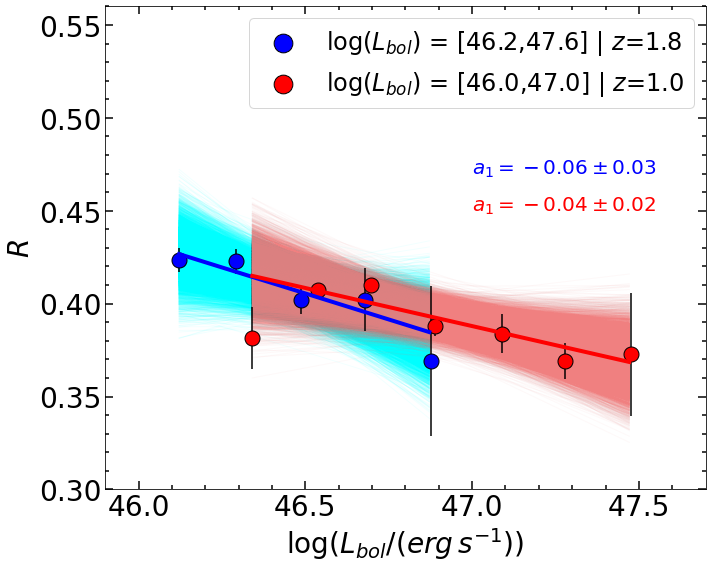}
\label{fig:ex_lbol_cf}
\end{subfigure}
\caption{\textit{Left:} $z$--$R$ (left) and $\log(L_{\rm bol})$--$R$ (right) section of the parameter space for the bins denoted by rectangles in Fig. \ref{fig:parspace_cf}. The solid lines represent the best linear fits to the data, while the shaded areas, obtained by resampling the posterior distributions of the best fit parameters, mark the 95\% confidence intervals.}
    
\label{fig:ex_correlations}
\end{figure*}

In Fig. \ref{fig:ex_correlations} we show, as an example, the $z-R$ and the $\log(L_{\rm bol})-R$ projections of the parameter space highlighted as rectangles in Fig. \ref{fig:parspace_cf}, together with the best-fit relations.

In general, the partial correlation coefficients clearly indicate $L_{\rm bol}$ as the main driver of the correlation ($\rho = -0.734$, p-value=$1 \times 10^{-10}$, $\tau = -0.537$, p-value=$<10^{-4}$), while at fixed luminosity $z$ plays no role in the evolution of $R$ ($\rho  = 0.048$, p-value=0.726, $\tau  = -0.099$, p-value=0.222).
The same finding is testified by the average slopes of the linear regressions: the anti-correlation between $L_{\rm bol}$ and $R$ is systematically detected at all redshifts with the average slope of the linear regressions being $\langle a_{1,L} \rangle = -0.051 \pm 0.012$. On the other hand, the average slope of the $z-R$ relation is on average consistent with zero within 1 $\sigma$, being $\langle a_{1,z} \rangle = -0.007 \pm 0.007$.

\begin{table}[h!]
\centering
\setlength{\tabcolsep}{5pt}
\begin{tabular}{c c c}
 \hline \noalign{\smallskip}
 Parameters & $\rho_{R \, x \mid y}$ (p-value) & $\tau_{R \, x \mid y}$ (p-value)\\
 (1) & (2) & (3) \\

 \hline \noalign{\smallskip}
 $z-R$ & $0.048$ ($0.726$) & $-0.099$ ($0.222$) \\ 
  
 $\log(L_{\rm bol})-R$ & $-0.734$ ($10^{-10}$) & $-0.537$ ($<10^{-4}$) \\

 \hline
\end{tabular}

\caption{Correlation parameters between $z$, $\log(L_{\rm bol})$ and $R$. Columns represent respectively: correlated parameters (1), partial correlation index between $R$ and parameter $x$ keeping $y$ fixed and associated p-value using the Spearman correlation index (2), partial correlation index between $R$ and parameter $x$ keeping $y$ fixed and associated p-value using the Kendall $\tau$ correlation index (3).}
\label{tbl:tbl_corr_1}
\end{table}

\begin{table}[h!]
\centering
\setlength{\tabcolsep}{5pt}
\begin{tabular}{c c c}
 \hline \noalign{\smallskip}
 Parameters & $\langle a_0 \rangle$ & $\langle a_1 \rangle$\\
 (1) & (2) & (3)\\

 \hline \noalign{\smallskip}
 $z-R$ &  $0.388 \pm 0.008$ & $-0.007 \pm 0.007$\\ 
  
 $\log(L_{\rm bol})-R$ & $0.392 \pm 0.004$ & $-0.051 \pm 0.012$\\

 \hline
\end{tabular}

\caption{Best fit parameters of the relations between $z$ , $\log(L_{\rm bol})$ and $R$. Columns represent respectively: fitted parameters (1) mean coefficients of the best-fit polynomial and relative standard errors of the means (2, 3).}
\label{tbl:tbl_corr_2}
\end{table}

These results suggest that the covering factor, here parametrised by it proxy $R$, does not evolve with redshift, and its evolution is only driven by the AGN luminosity. A direct consequence of this finding, on which we will come back in the discussion, is the self similarity of the innermost region of the AGN throughout all cosmic times, at fixed luminosity.

\subsection{The full sample SED}
\label{sec:seds}
\begin{figure}[h!]
\centering
\includegraphics[width=\linewidth,clip]{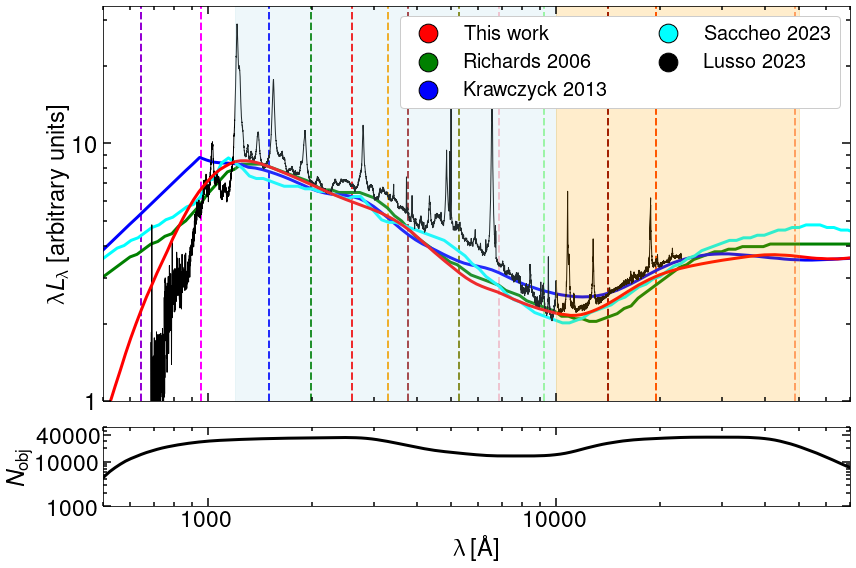}
\caption{Our full sample SED and others from literature (see text). The vertical dashed lines represent the effective wavelengths of the photometric filters used in this work, adopting the same colour code as in Fig. \ref{fig:corrections} blue-shifted to the mean redshift of the sample. All the SEDs are normalized by their integrated luminosity between 1,000 \AA\, and 1 $\mu$m. The AGN spectral composite from \citet{lusso2023euclid} is scaled to match our SED (red solid line) at 2,200 \AA. The black line in the bottom panel shows the number of sources contributing per spectral channel in the average SED. The full-sample composite SED is available in its entirety in a machine-readable form in the online journal.}
\label{fig:sed_comparison}
\end{figure}

In Fig. \ref{fig:sed_comparison} we show the comparison between the total composite obtained using our full sample with other SED templates in the literature (\citealt{richards2006spectral, krawczyk2013mean, saccheo2023wissh}). Additionally, we show in black the spectral composite from \citet{lusso2023euclid} to compare our photometry to Euclid NIR spectroscopy. The SEDs are scaled by their integrated luminosity between 1,000 \AA\, and 1 $\mu$m for a proper comparison. In the optical, our global SED resembles those from \citet{richards2006spectral} and \citet{krawczyk2013mean}, as expected since also these works were based on SDSS samples. In the IR the WISSH average SED from \citealt{saccheo2023wissh} outshines the SDSS-based samples, again unsurprisingly, being the WISSH quasars IR-selected. The small blue bump associated with the \ion{Fe}{ii} UV pseudo-continuum between 2250--3300 \AA\ (e.g. \citealt{grandi19823000}), visible in the other average SEDs, instead appears to have been well subtracted in ours. 


\subsection{Evolution of $CF$ with the Eddington ratio}
\label{sec:er_evolution}
The Eddington ratio expresses the efficiency of the accretion process in AGN, with sources emitting close or above the Eddington ratio having reached their maximal radiative emission under the assumption of spherical accretion. Despite this simplistic approximation, this parameter represents a reasonable proxy for the normalised accretion rate. By scaling the emitted luminosity by the black hole mass (which sets the black hole sphere of influence), the Eddington ratio is a \textit{scale invariant} quantity, in contrast with the bare $L_{\rm bol}$. Tori with similar covering factors possibly share similar accretion properties (see e.g. \citealt{ricci2023bass} and references therein).

\begin{figure}[h!]
\centering
\includegraphics[width=\linewidth,clip]{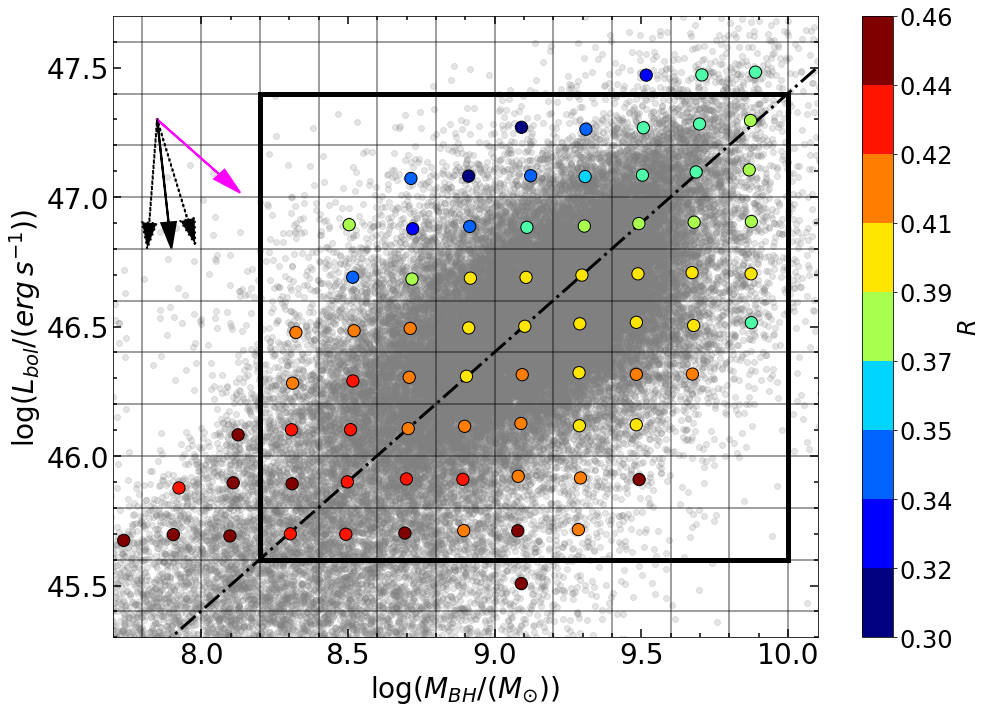}
\caption{$\log(M_{\rm BH})$--$\log(L_{\rm bol})$ plane with $R$ in colour-code. The black rectangle shows the region where the PCA was evaluated. The black solid and dotted arrows represent the direction of the strongest correlation as in Fig. \ref{fig:parspace_cf}. The dot-dashed line marks a constant $\lambda_{\rm Edd}$=0.2, representing the average value of the sample, while the magenta arrow shows the direction of decreasing $\lambda_{\rm Edd}$.}
\label{fig:lm_parspace_R}
\end{figure}

With the aim of investigating this possibility, we explored the evolution of the covering factor within the accretion parameter space, i.e. the parameter space given by $\log(M_{\rm BH})$ and $\log(L_{\rm bol})$\footnote{Although the axes of this parameter space are physically not correlated, we highlight that from an observational standpoint they are. Both the estimators for $M_{\rm BH}$ -computed via the virial calibrations- and $L_{\rm bol}$ -obtained through bolometric corrections- depend generally on the same continuum monochromatic luminosity.}. Since we just showed that the seeming dependence of the covering factor on $z$ actually derives from the effect of the luminosity, here we do not take into account any redshift effect. For this analysis, we basically repeated the same steps adopted to build the $\log(L_{\rm bol})$-$z$ parameter space, keeping the same binning width. We then proceeded to produce the average SEDs and evaluate $R$ accordingly. We also computed in each bin the average $\lambda_{\rm Edd}$. This alternative parameter space, together with the $R$ parameter in colour-code, is shown in Fig. \ref{fig:lm_parspace_R}. Considering the values of the $\log(L_{\rm bol})$--$R$ and the $\rm \log(\lambda_{\rm Edd})$--$R$ standard correlations (i.e. non partial) we find, as expected, significant correlations for both the relations. In particular, for $\log(L_{\rm bol})$--$R$ we report $\rho$=$-$0.895 with p-value$<10^{-20}$ and for $\log(\lambda_{\rm Edd})$--$R$ $\rho$=$-$0.465 with p-value $10^{-4}$. Although both anti-correlations are statistically significant, the face values suggest that the stronger driver of the two, in the evolution of the covering factor, is the luminosity. Similar conclusions were reached by employing the $\tau$ index of correlation, which gives $\tau$=$-$0.708 (p-value = $8\times 10^{-17}$) and $\tau$=$-$0.346 (p-value = $5\times 10^{-5}$) respectively for $L_{\rm bol}$--$R$ and $\lambda_{\rm Edd}$--$R$. A more informative way to tackle this question is to perform again a PCA, this time within the accretion parameter space, and check \textit{a posteriori} the direction of the strongest correlation. We performed this test on the bins residing in the region quite uniformly populated, marked as a black rectangle in Fig. \ref{fig:lm_parspace_R}. As described in Sec. \ref{sec:lbol_correlation}, we evaluated the PCA between $\log(L_{\rm bol})$ and $R$, while keeping $\log(M_{\rm BH})$ fixed and vice-versa. We also evaluated the uncertainty on the direction of the strongest correlation by performing a bootstrap resampling. The PCA arrow aligns almost completely ($\theta =-84^{\circ}$) with the direction of decreasing $\log(L_{\rm bol})$. In Fig. \ref{fig:lm_parspace_R} we show the PCA direction of increasing $R$ as a black arrow. As a comparison, we also mark as a magenta arrow the direction of decreasing $\lambda_{\rm Edd}$, which should be followed by an increase of the covering factor if the Eddington ratio were the main driver of the $CF$ evolution. Again, $L_{\rm bol}$ appears as the ultimate driver in the evolution of $CF$.

Again, $L_{\rm bol}$ appears as the main driver in the evolution of $CF$, despite the large systematic uncertainties on the black hole masses of the current calibrations (0.3--0.5 dex, see e.g. \citealt{vestergaard2006determining}), which do not provide reliable estimates of $\lambda_{\rm Edd}$, thus possibly diluting any correlation with $\lambda_{\rm Edd}$.

\subsection{The non-evolution with $z$ of the torus in AGN: implications for the X-ray obscured fraction}

The analyses carried out so far strongly suggest a non-evolution with $z$ of the torus properties, at least in its innermost region for the blue type 1 AGN included in this sample. When this evidence is coupled with the seeming non-evolution of the optical/UV SED between local and high redshift quasars, it appears that the nuclear structure in blue quasars only depends on the (mass-normalised) luminosity of the nucleus, regardless of the cosmic epoch. 
Within the Unified Model, a redshift-independent torus geometry translates into a constant fraction of optical type 1 AGN at fixed luminosity (see e.g. Fig. 6 in \citealt{merloni2003}) at all cosmic times. As an application of this result, we can employ the covering factor derived from $L_{\rm NIR}/L_{\rm opt}$ to set a lower limit for the X-ray obscuration. X-ray obscuration can be indeed delivered on different scales (nuclear/circumnuclear and galactic), and by dusty and/or dust-free gas. If the amount of obscuration  imparted by the torus does not vary with redshift, any kind of evolution in the obscuring medium should occur elsewhere.

We thus aimed at comparing the covering factor here derived via the $L_{\rm NIR}/L_{\rm opt}$ ratio with that estimated in several X-ray surveys. We achieved this by converting our proxy into the actual covering factor adopting standard prescriptions. In particular, we show that, adopting reasonable assumptions regarding the disc emission and the the torus opening angle, it is possible to reproduce the X-ray obscured fraction at $z\sim$0, and set a lower limit for the X-ray obscuration.



To assess the degree of obscuration ascribable to the torus and compare it with other samples in literature, we selected all the bins in the $\log(L_{\rm bol})-z$ plane between $46.8 \leq \log (L_{\rm bol}) \leq 47.2$ and $0.9\leq z \leq 2.9$. This region of the parameter space is densely populated and allows us to sample a wide redshift interval also covered by the X-ray samples. We then converted the observed $R=L_{\rm NIR}/L_{\rm opt}$ into the covering factor $CF$ by scaling these observed quantities respectively to the total torus and disc luminosities following this procedure. First, we assumed that the disc emission comes from a standard optically thick and geometrically thin accretion disc (\citealt{shakura1973black}). We evaluated its SED in each bin using the average $M_{\rm BH}$ and converting the observed $L_{\rm bol}$ into the accretion rate assuming a standard mass-to-luminosity conversion efficiency of 0.1. We also explored other values for the efficiency, but the variations in the plausible range of opening angles were of the order of some degrees. We then scaled the far infrared quasar template from \citet{lyu2017intrinsic} to match the 1-$\mu$m emission of the accretion disc. This allowed us to take into account also the FIR torus emission up to 1,000 $\mu$m. Lastly, we computed the total disc luminosity between 10 \AA\, and 1 $\mu$m and the total torus luminosity between 1 $\mu$m and 1,000 $\mu$m, and used these quantities to scale the observed luminosities which are instead integrated between $\sim$1,000 \AA--1 $\mu$m, and 1--5$\mu$m. In order to explore the effect of the anisotropy of the radiation field of the accretion disc and the clumpiness of the torus, we also employed the total disc and torus luminosities to evaluate the $CF$ adopting the \citet{stalevski2016dust} prescriptions. 

\begin{figure}[h!]
\centering
\includegraphics[width=\linewidth,clip]{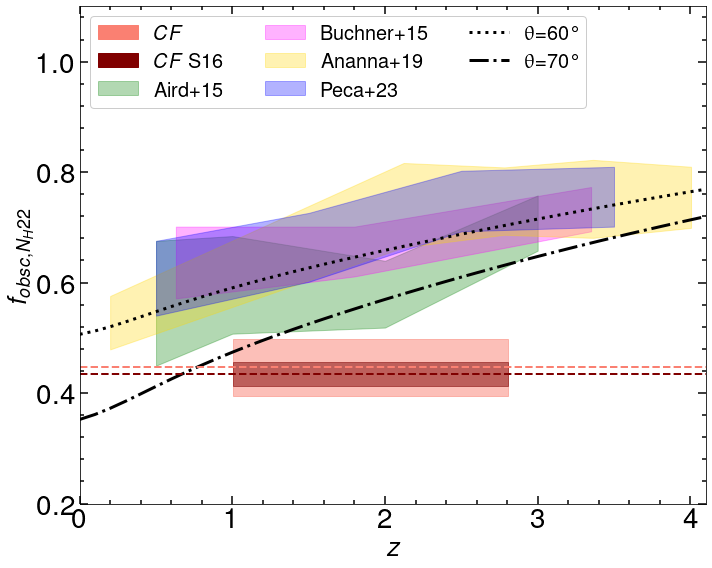}
\caption{Obscured AGN fraction with $N_{\rm H}>10^{22}$ cm$^{-2}$ as a function of redshift in different literature samples (\citealt{aird2015evolution, buchner2015obscuration, ananna2019accretion, peca2023cosmic}, including a correction for Compton thick sources, see text for details. The horizontal shaded areas represent the different inferred ratios between the torus and the disc luminosities, as well as the intrinsic $CF$ according to the \citet{stalevski2016dust} prescriptions. These values are extrapolated at the redshifts not directly covered by our bins at $\log(L_{\rm bol})\sim 47$ accordingly to the dashed lines. The black dotted and dot-dashed lines correspond respectively to the \citet{gilli2022supermassive} models for $\delta$=3.3, $\gamma$=2.0 and $\theta$=60$^{\circ}$, and 70$^{\circ}$.}
\label{fig:fobsc_z}
\end{figure}

We show the result of this approach in Fig. \ref{fig:fobsc_z}. There, we compare the $CF$ values derived here to the fraction of X-ray obscured objects (with column density $N_{\rm H} > 10^{22}$ cm$^{-2}$) according to different literature samples (\citealt{aird2015evolution, buchner2015obscuration, peca2023cosmic, ananna2019accretion}). When the AGN sample included only Compton-thin objects, we adjusted the obscured AGN fractions by assuming an equal number of Compton-thin and Compton-thick sources (e.g. \citealt{gilli2007synthesis, ananna2019accretion}. We chose the X-ray luminosity regime $L_{X}\sim 10^{45}$ erg s$^{-1}$, which roughly corresponds to $L_{\rm bol}\sim 10^{47}$ erg s$^{-1}$ adopting the \citet{duras2020universal} X-ray to bolometric conversion. Here the obscured fractions of the reference samples were corrected for the effect of the Malmquist bias, which reduces the fraction of observed obscured sources, by assuming intrinsically equal amounts of obscured an unobscured sources (i.e. p=0.5 in Eq. 1 in \citealt{signorini2023x}). 

In the plot are also highlighted the curves from the analytical model described in \citet{gilli2022supermassive} accounting for both the ISM and the torus absorption in the X-rays, corresponding to $\delta=3.3$, $\gamma=2.0$, and $\theta=60^\circ$ (black solid line), and $\theta=70^\circ$ (black dotted line), which encompass the possible range of $CF$ values. Here $\delta$ is the exponent in the relation between the column density ($N_{\rm H}$) and $z$, i.e. $N_{\rm H} \propto (1+z)^{\delta}$, $\gamma$ is the exponent in the relation describing the evolution of the surface density of the molecular clouds ($\Sigma$) with $z$, that is $\Sigma \propto (1+z)^{\gamma}$ and $\theta$ is the torus half-opening angle evaluated from its vertical axis. We also display as shaded rectangles the inferred $CF$ directly estimated as the infrared-to-optical luminosity ratio (salmon) as well as according to the \citet{stalevski2016dust} recipe (maroon). The lines guide the eye at the redshifts not actually sampled by the $L_{\rm bol}\sim 10^{47}$ erg s$^{-1}$ bins.

We find that, assuming a plausible range of opening angles, the model curves manage to reproduce the X-ray obscured fraction, which at $z=0$ can be explained by the torus alone. The resulting opening angles implied by the model are consistent with other independent estimates derived from modelling of the infrared emission in type 1 AGN (e.g. \citealt{almeida2011testing, ichikawa2015differences}). Interestingly, also studies not directly aimed at investigating the torus properties provided evidence for similar values. For instance, \citet{signorini2024quasars}, while attempting to explain the remarkably small intrinsic dispersion of the X-ray to UV luminosity relation in quasars, found -- in a completely independent way -- the need for the presence of a toroidal structure obscuring the accretion disc with an opening angle $\gtrsim$65$^{\circ}$ from the disc axis.

Although the obscuration provided by the torus does not appear to vary with $z$, the X-ray obscured fraction does instead appear to increase. This observational trend calls for some sort of redshift-dependent absorber. Yet, the nature of this additional component is not clear, nor are the scales where the absorption of the X-rays should happen. In principle, an increase of dust-free gas in the micro-scale (i.e. the BLR) and/or dust and gas in the macro-scale (i.e. the ISM scale) towards earlier cosmic times could go in the direction of bridging between the degree of obscuration derived in optical/infrared and X-ray surveys.

\section{Discussion}
\label{sec:discussion}

Understanding the evolution of the covering factor and its dependence on the parameters regulating the accretion process is a crucial step towards unveiling the complicated and dynamic structure of AGN. In this effort, valuable information can be gathered by exploring the broadband correlations between the main emission from the accretion disc and the reprocessed emission by the dusty torus in the infrared. The large photometric dataset presented here, spanning roughly two and a half orders of magnitude in terms of bolometric luminosity, has been purposefully assembled to explore these correlations up to $z\sim 3$, when the Universe was about 2 Gyr old.

Here, we showed that the disc luminosity acts as the main driver in the evolution of the NIR properties in our sample, representative of blue unobscured type 1 AGN. This results in an anti-correlation between the AGN luminosity and the torus covering factor, here parameterised via the ratio $R$ between the NIR and the UV/optical luminosity. Conversely, the redshift does not play a significant role, as the $CF$ does not evolve between low- and high-$z$ objects at fixed luminosity.

The large sample analysed in this work provides compelling evidence against a redshift evolution of the dusty torus geometry, at least in its innermost, hottest region. The NIR SED, once we control for the dependence on the luminosity and take into account the effect of the shifting wavebands, does not exhibit any evolution with $z$. The most likely driver in the covering factor appears to be the AGN luminosity, whose intensity directly shapes the circumnuclear surroundings. In a broader perspective, the similarity of the SEDs between low- and high-redshift quasars is remarkable. The optical/UV SED of blue unobscured AGN are barely sensitive to the passing of cosmic time since $z\sim$7 (e.g. \citealt{kuhn2001search, mortlock2011luminous, hao2013spectral, shen2019gemini, yang2021probing, trefoloni2024quasars}). Similar considerations apply to the interplay between the accretion disc and the X-ray coronal emission. Their coupling, epitomised in the $L_{\rm X}-L_{\rm UV}$ relation (\citealt{avnitananbaum79}), does not exhibit any appreciable redshift evolution up to $z\sim 7$ (\citealt{lusso2015, risaliti2019cosmological, salvestrini2019, nardini2019, lusso2020}) once observational biases have been removed. 
The similarity between low- and high-redshift quasars also holds in terms of broad emission line properties. Several works investigating the evolution of the broad lines in large samples of quasars have indeed demonstrated the absence of significant redshift trends at fixed luminosity (\citealt{croom2002correlation, nagao2006, stepney2023no}). Analogue results have been obtained when investigating the chemical enrichment of the BLR, parameterised, for instance, via the UV \ion{Fe}{ii} to \ion{Mg}{ii}$\lambda$2798 ratio, which does not evolve up to $z\sim 7$ (\citealt{dietrich2003fe, mazzucchelli2017physical, sameshima2020mg, wang2022metallicity, trefoloni2023most, jiang2024no}).
Putting together these pieces of evidence, what follows is that the quasar nuclear environment, i.e. the ensemble made of corona, accretion disc, BLR, and torus, is barely sensitive to the redshift, once the accretion parameters are fixed. The striking similarity between high- and low-redshift AGN goes in the direction of an early assembly of the nuclear region in luminous AGN, with the BLR of high-redshift quasars being already enriched by $z\sim7$ and the circumnuclear torus shaped according to the quasar radiation field.

A direct comparison with other works exploring the correlations between the $CF$ and the accretion parameters is not straightforward. There are indeed several technical limitations on the available wavebands at different redshifts, as well as different proxies adopted in literature to estimate the covering factor. For instance, earlier works on the PG quasars (\citealt{cao2005dust}) adopted the 3--10 $\mu$m interval to estimate $L_{\rm IR}$, while both \cite{gu2013evolution} and \citet{ralowski2023covering} adopted the 1--7 $\mu$m interval and, more recently, \citet{wu2023ensemble} took advantage of the 1--10 $\mu$m band. Conversely, \citet{lusso2013obscured} and \citet{ichikawa2019bat} performed a torus--galaxy decomposition first, and then integrated the best-fit torus model between 1--1,000 $\mu$m. Similar considerations apply to the choice of $L_{\rm opt}$. Despite these minor differences in the adopted proxies to the covering factor, our results follow the trend generally observed in studies using $L_{\rm IR}/L_{\rm opt}$ as a proxy to $CF$, which find an anti-correlation between $CF$ and both the disc luminosity and the Eddington ratio (\citealt{treister2008measuring, maiolino2007dust, calderone2012wide, ma2013covering, ezhikode2017determining, toba2021does, ralowski2023covering}, although there are also claims of non-correlations e.g. \citealt{gu2013evolution, zhuang2018infrared}). It is also important to remark that the choice of the IR integration interval does not only play a role in the comparison between different literature works, but is also key when it comes to assessing the strength of the anti-correlation between the disc and the torus luminosities. For example, when the MIR wavelengths ($\gtrsim$ 10 $\mu$m) are included in the torus luminosity  estimate, the cooler polar dust emission can add up to that coming from the torus, hence diluting the correlation (\citealt{wu2023ensemble}).

An update of the canonical unification paradigm, the so-called ``radiation-regulated unification model'' (\citealt{ricci2017}), adds the accretion rate as a further parameter playing a crucial role in the classification of an AGN. In this framework, highly accreting sources are more efficient in clearing their close environment and shaping the circumnuclear toroidal structure, by means of radiation pressure (\citealt{pier1992infrared, fabian2008effect}) and outflows (\citealt{fabian2009radiation}). The correlation observed between $CF$ and $\lambda_{\rm Edd}$ in our data supports this scenario. Yet, the degree of correlation between $R$--$L_{\rm bol}$ and $R$--$\lambda_{\rm Edd}$ favours the former as a stronger driver of the $CF$ evolution. There are several explanations for this result: if the luminosity were the actual stronger driver, expressing the correlation in terms of observed quantities for the bulk of the sample, $L_{\rm bol} \propto L_{3000\AA}$ and $\lambda_{\rm Edd} \propto L_{3000\AA} / (\rm FWHM_{MgII}^{2} \, L_{3000\AA}^{1/2}) \propto L_{3000\AA}^{1/2}$, the lower degree of correlation would come from the lower exponent of the luminosity term. Conversely, it is also possible that the Eddington ratio is the actual parameter driving the evolution of the $CF$, but the large systematic uncertainties on the $M_{\rm BH}$ estimates reduce the degree of correlation. Lastly, it is also possible that the Eddington ratio is actually an important factor in shaping the dusty torus, but not in the $\lambda_{Edd}$ range covered by our sample. This is what has been recently shown in recent studies, parameterising the $CF$ as the fraction of X-ray obscured sources. These works argued that the strongest decrease in the $CF$ for increasing $\lambda_{\rm Edd}$ occurs between $\log(\lambda_{\rm Edd})\simeq$ $-3$ and $-1$ (\citealt{ricci2017, mizukoshi2024updated}). As a comparison, the bulk of our sample has $\lambda_{\rm Edd}\gtrsim-1.2$.


Once the redshift evolution of the torus covering factor is abandoned in favour of the accretion parameters being the sole drivers in the torus shape, any redshift-dependent obscuration, such that reported in X-ray surveys, must be delivered elsewhere. Generally speaking, X-ray obscuration can occur on different scales, ranging from nuclear to galactic scales. Although there is strong evidence in favour of X-ray obscuration happening also because of dust-free gas within the BLR (\citealt{maiolino2001dust, risaliti2002ubiquitous, goulding2012deep, ichikawa2019bat, ricci2022bass}), there are not observational clues about gas richer BLR with increasing redshift: as already mentioned, luminosity-matched samples of Type 1 AGN at low and high redshift display on average broad emission lines with comparable strength (e.g. \citealt{croom2002correlation, stepney2023no}).
However, on galactic scales, the combination of increasing ISM mass (e.g. \citealt{scoville2017evolution}) and decreasing galactic sizes with redshift (e.g. \citealt{allen2017size}) naturally implies the increase in the ISM density. This, in turn, could provide an additional source of X-ray opacity (\citealt{gilli2022supermassive, alonso2024probing}), whose relevance is expected to increase with $z$, thus going in the direction of increasing the fraction of X-ray obscured sources.

\section{Conclusions}
\label{sec:conclusions}

We gathered a large photometric dataset containing $\sim$36,000 blue quasars to analyse separately the evolution of the NIR properties of blue Type 1 AGN with the redshift and the bolometric luminosity.
Our main findings are listed below:

\begin{itemize}
    
    \item The average NIR emission in different luminosity bins clearly shows a decreasing trend for increasing $L_{\rm bol}$. Once we take into account the effect of the shift of the sampling wavebands at different redshifts, the average NIR SEDs at different $z$ are mostly consistent with an intrinsic standard torus SED. This finding, coupled with the other pieces of evidence of non evolution of accretion disc, BLR and coronal emission in blue quasars, goes in the direction of a universal (but luminosity-dependent) quasar SED.
    
    \item All the explored proxies for the covering factor do not correlate with the redshift, once the bolometric luminosity is kept fixed.

    \item  the ratio, $R$, between the NIR and optical/UV luminosities, our proxy to the intrinsic $CF$, anti-correlates with the bolometric luminosity ($\rho = -0.734$) regardless of the redshift. This finding fits within the so-called ``receding torus'' scenario, where the radiation field from the inner AGN is capable of shaping the structure of the dusty torus.
        
    \item $R$ also significantly anti-correlates with the Eddington ratio, but this trend is likely due to the dependence of $\lambda_{\rm Edd}$ on $L_{\rm bol}$. This is confirmed by the fact that the direction of strongest variation of $R$ in the accretion parameter space aligns with $L_{\rm bol}$ rather than with $\lambda_{\rm Edd}$.

    \item A direct consequence of the non evolution of the torus covering factor with $z$ is that any increase in the X-ray obscuration with redshift, if systematically detected, should occur by means of some other mechanism, either on nuclear or galactic scales.

\end{itemize}

Our systematic investigation of the interplay between the optical and infrared broadband properties of type 1 AGN provides evidence for a standard NIR torus SED. The properties of the torus seem to be mostly governed by the luminosity of the nucleus, regardless of the redshift.
In spite of the wide sample gathered here, a key step forward will be the detection of a statistically significant number of faint high-redshift objects in the rest-frame infrared, to test whether this behaviour also holds in the low-luminosity tail of the AGN population at high $z$. As far-infrared missions such as the Origins Space Telescope (\citealt{battersby2018origins}) still stand out in a not so near future, most of the efforts should be directed in two main directions. A thorough theoretical understanding of the interplay between the disc emission and the patchy torus, taking into account also the action of a disc wind in the process of shaping the circumnuclear environment (e.g. \citealt{chan2016radiation} and references therein), would prove of utmost importance. 
At the same time, a panchromatic effort involving both new and archival X-ray data would definitively deepen our knowledge in the canonical classifications between optical and X-ray type 1/2 AGN.

\begin{acknowledgements}
We acknowledge financial support by the INAF Grants for Fundamental Research 2023.
For catalogue cross analysis we have used the Virtual Observatory software \topcat (\citealt{taylor2005_topcat}), available online at http://www.star.bris.ac.uk/mbt/topcat/. The figures were generated with \matplotlib (\citealt{hunter2007matplotlib}), a PYTHON library for publication-quality graphics.
\end{acknowledgements}

\bibliographystyle{aa} 
\bibliography{bibl}

\begin{appendix}

\twocolumn

\section{On the reliability of WISE photometry}
\label{app:wise_phot}
During its primary mission WISE measured the full sky in four mid infrared bands centered on 3.4, 4.6, 12, and 22 $\mu$m, but the exhaustion of the solid hydrogen cryogen made W4 unusable. After this, the survey continued another half-sky scan in W1, W2, and W3. For the sake of comparison, before the reactivation of the WISE survey (NEOWISE, \citealt{mainzer2014initial}), the median coverage in the W1 and W2 bands was 33 exposures, 24 in W3 and only 16 in W4 (\citealt{lang2016wise}). This, combined with the lower relative response made the W4 WISE survey significantly shallower than in the other bands.

\begin{figure}[h!]
\centering
\includegraphics[width=\linewidth,clip]{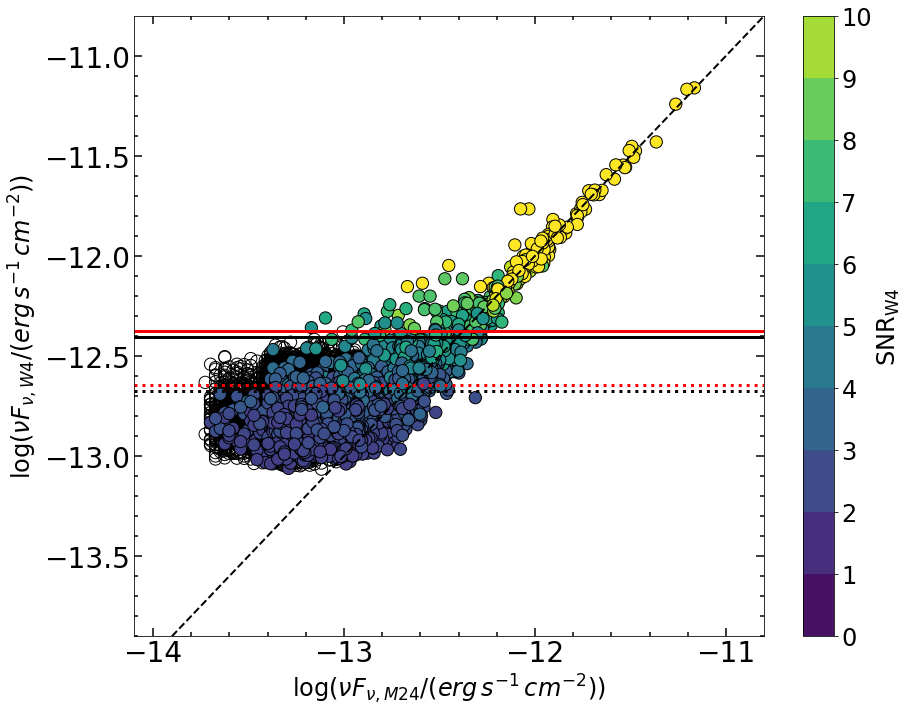}
\caption{Comparison between SPITZER MIPS24 and W4 photometry as a function of W4 SNR in colour-code. Empty circles in the background denote upper limits. W4 measurements are systematically overestimating the actual fluxes at low SNR. The black solid (dotted) horizontal line marks the median (lowest 5$^{th}$ percentile) flux for the objects in our sample without performing the quality cuts described in the text. The same holds for the red lines where also the SNR$\geq$2 cut is applied. A significant amount of our sources would have systematically overestimated W4 fluxes.}
\label{fig:w4_m22}
\end{figure}

The inclusion of these photometric data could, in theory, bias towards higher than the average IR fluxes, especially at high redshift ($z>2$) where the 1--5 $\mu$m range becomes increasingly contained within the W4 passband. In order to check whether this is the case, we compared the Spitzer First Look Field (FLS, PID: 26; PI: B. T. Soifer) observed by the SPITZER Multiband Imaging Photometer for SPITZER (MIPS) photometry at 24 $\mu$m (M24) with that of WISE W4.
To this purpose, we cross-matched the SPITZER FLS MIPS24 (16,905 sources) and the AllWISE point source catalogues adopting a 2'' matching radius, obtaining 11,253 matches. In Fig. \ref{fig:w4_m22} we show the result of the comparison, colour-coding the W4 SNR. It is evident that at low fluxes (and SNR) the W4 photometry, including also upper limits which amount to $\sim$75\% of the matches, overestimates the actual flux. The high incidence of the upper limits is likely due to the low sensitivity in the W4 band, while the large fraction of overestimations likely resulting from the larger aperture (16.5'') adopted for the W4 filter. Because of this systematic overestimation at low fluxes, we refrained from using W4 photometry. We show as solid and dotted lines the median and lowest 5$^{th}$ percentile of the flux distribution for our sample adopting the quality cuts described in the main text, i.e. mag$i$$<$19.0, the colour selection and the choice of the 0.1$\leq z \leq 2.9$ redshift interval. The red lines represent the same values after adding also the SNR$\geq$2 criterion on W3 photometry.

\begin{figure}[h!]
\centering
\includegraphics[width=\linewidth,clip]{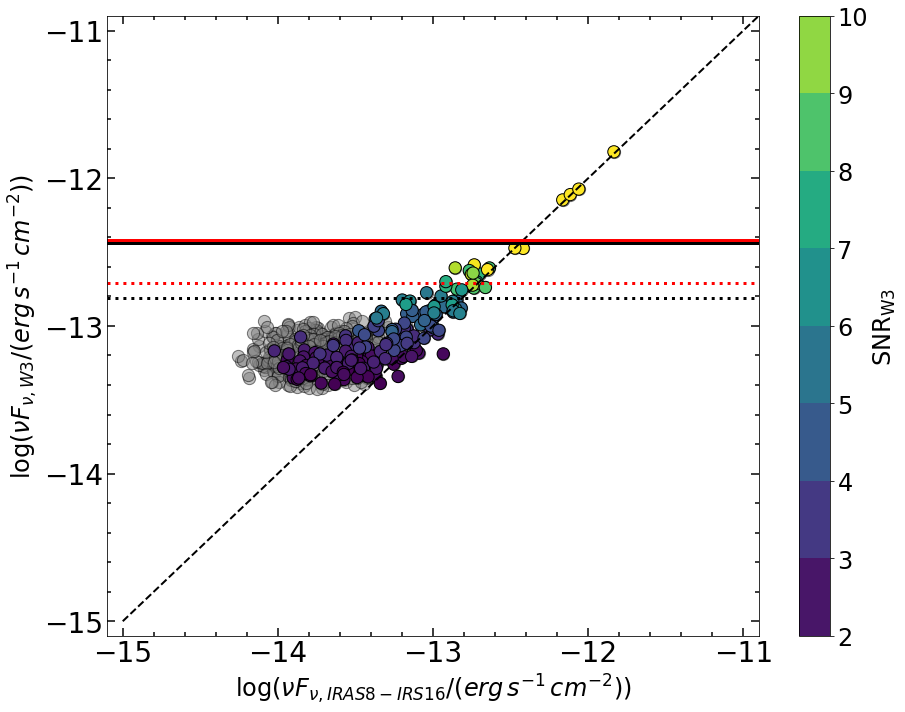}
\caption{Comparison between the 12-$\mu$m fluxes interpolated between SPITZER IRAC 8 $\mu$m and IRS 16 $\mu$m and W3, with the W3 SNR in colour-code. The grey dots denote sources below the W3 SNR=2 threshold chosen in this work. The black solid (dotted) horizontal line marks the median (lowest 5$^{th}$ percentile) flux for the objects in our sample without performing the quality cuts described in the text. The same holds for the red lines where also the SNR$\geq$2 cut is applied.
Although there is a trail of points detaching from the one-to-one relation at low SNR W3 ($\sim$4--5), roughly 95\% of our sample has fluxes higher than the threshold where W3 starts to systematically overestimate the Spitzer measurements. The colour-code of the lines is the same as in Fig.\ref{fig:w4_m22}.}
\label{fig:w3_i8_16}
\end{figure}

We also performed the same exercise to check the reliability of the W3 measurements. However, we only managed to perform this check on a much smaller (557 sources) sample of targets, obtained by cross-matching the Spitzer/IRAC and Spitzer/IRS data in the GOODS North and South fields (\citealt{giavalisco2004great}) with the AllWISE catalogue. In order to evaluate the monochromatic flux at the W3 reference wavelength (12 $\mu$m), we logarithmically interpolated the flux of the two closest Spitzer bands, namely IRAC 8 $\mu$m and IRS 16 $\mu$m. We show the result of this match in Fig. \ref{fig:w3_i8_16}. Also in this case we observe a trail detaching from the one-to-one relation, although the average difference between the $\log F_{\nu}$ for Spitzer and WISE is lower than in the W4 vs. MIPS24 comparison ($-$0.22 dex after selecting sources with SNR W3 $\geq$2 against $-$0.42). Also in this case we observe the lowest SNR points being increasingly more offset from the identity line. Yet, in this case even the lowest 5$^{th}$ percentile of the flux distribution for the SDSS-WISE sample lies above the flux value where W3 starts to systematically overestimate the actual flux with respect to Spitzer. This implies that, within our sample, the effect of overestimating the W3 fluxes has a negligible impact.

\begin{table*}
\centering
\setlength{\tabcolsep}{2pt}
\begin{tabular}{c c c c c c}
 \hline \noalign{\smallskip}
 Subset & Parameters & $\rho_{R \, x \mid y}$ (p-value) & $\tau_{R \, x \mid y} $ (p-value) & $\langle a_0 \rangle$ & $\langle a_1 \rangle$\\
 (1) & (2) & (3) & (4) & (5) & (6)\\

 \hline \noalign{\smallskip}
 $R_{2-5 \mu m}$ & $z-R$                   & $0.459$ ($4\times 10^{-4}$) & $0.133$ ($0.100$) & $0.247 \pm 0.003$ & $0.009 \pm 0.005$\\ 
 $R_{2-5 \mu m}$ & $\log(L_{\rm bol})-R$   & $-0.737$ ($10^{-10}$) & $-0.465$ ($<10^{4}$) & $0.250 \pm 0.001$ & $ -0.031 \pm 0.009$\\
 \hline
 
 $R_{1-5 \mu m \, peak}$ & $z-R$                   & $-0.028$ ($0.836$) & $-0.104$ ($0.196$) & $0.425 \pm 0.013$ & $-0.003 \pm 0.017$\\ 
 $R_{1-5 \mu m \, peak}$ & $\log(L_{\rm bol})-R$   & $-0.448$ ($6\times10^{-4}$) & $-0.328$ ($<10^{4}$) & $0.424 \pm 0.007$ & $ -0.074 \pm 0.019$\\
 \hline
 
 $R_{S16}$ & $z-R$                         & $0.048$ ($0.726$) & $-0.099$ ($0.219$) & $0.580 \pm 0.005$ & $-0.005 \pm 0.004$\\ 
 $R_{S16}$ & $\log(L_{\rm bol})-R$         & $-0.740$ ($10^{-10}$) & $-0.537$ ($<10^{4}$) & $0.583 \pm 0.002$ & $ -0.031 \pm 0.008$\\
 \hline
 
 $CF_{L13}$ & $z-R$                        & $0.048$ ($0.726$) & $-0.099$ ($0.222$) & $0.279 \pm 0.004$ & $-0.004 \pm 0.003$\\ 
 $CF_{L13}$ & $\log(L_{\rm bol})-R$        & $-0.740$ ($10^{-10}$) & $-0.537$ ($<10^{4}$) & $0.282 \pm 0.002$ & $ -0.026 \pm 0.006$\\\\
 \hline
 
 no-corr & $z-R$                           & $0.155$ ($0.260$) & $-0.0459$ ($0.572$) & $0.331 \pm 0.006$ & $-0.002 \pm 0.005$\\ 
 no-corr & $\log(L_{\rm bol})-R$           & $-0.746$ ($7\times 10^{-11}$) & $-0.533$ ($<10^{4}$) & $0.334 \pm 0.002$ & $ -0.043 \pm 0.011$\\
 \hline
 
 no-col & $z-R$                            & $0.011$ ($0.937$) & $-0.142$ ($0.077$) & $0.371 \pm 0.008$ & $-0.011 \pm 0.010$\\ 
 no-col & $\log(L_{\rm bol})-R$            & $-0.726$ ($4 \times 10^{-10}$) & $-0.531$ ($<10^{4}$) & $0.375 \pm 0.004$ & $ -0.055 \pm 0.013$\\
 \hline
\end{tabular}

\caption{Correlation parameters between $z$ , $\log(L_{\rm bol})$ and alternative $CF$ proxies also considering alternative data sets. Columns represent respectively: alternative $CF$ proxy or alternative data-set (1) described in \ref{app:checks}, correlated/fitted parameters (2), Spearman partial correlation index between $CF$ and parameter $x$ keeping $y$ fixed and relative p-value (3). Kendall $\tau$ partial correlation index between $CF$ and parameter $x$ keeping $y$ fixed and relative p-value (4). Mean coefficients of the best-fit polynomial and relative uncertainties (5, 6).}
\label{tbl:tbl_altcorr}
\end{table*}

\section{Consistency checks}
\label{app:checks}
The selection of 5 $\mu$m as the longest wavelength to sample the torus emission was chiefly dictated by the trade-off between choosing a wavelength well within the IR regime and the maximal observability throughout the sample. At the same time, the choice of employing the ratio $L_{\rm NIR}/L_{\rm opt}$, evaluated respectively between 1--5 $\mu$m and 1,000 \AA\,-- 1 $\mu$m, as a proxy for the intrinsic $CF$ was mainly due to our data-driven approach. As consistency checks, we performed the same analyses described in the main text adopting different proxies for the NIR term of the $CF$ and alternative data sets. 
\begin{itemize}
    \item We explored the adoption of the 2--5 $\mu$m interval for estimating the IR term contained in $R$ (subset $R_{2-5\,\mu m}$ in Table \ref{tbl:tbl_altcorr}). A longer wavelength for the lower end of the IR term is more reliable when it comes to estimate the torus emission, being less prone to stellar emission within the host galaxy. The price for this choice is, however, the narrowing of the integration interval and the consequent lowering of the $R$ value. 

    \item We explored the possibility of integrating the accretion disc SED between its peak and 1 $\mu$m rather than in the fixed 1,000 \AA\,--1 $\mu$m interval (subset $R_{1-5 \mu m \, \rm peak}$).
        
    \item With the aim of taking into account the effect of the anisotropy of the accretion disc radiation field, as well as the clumpiness of the torus, we tested the parameterisation described in \citealt{stalevski2016dust}. To this end, we scaled $R$ to the total ratio between the 1--1,000 $\mu$m infrared luminosity and the disc luminosity. We achieved this by adopting the quasar template SED built in Sec.\ref{sec:nir_evolution} (subset $CF_{S16}$).
    
    \item We also explored the $CF$ the expression described in \citet{lusso2013obscured}, i.e. $CF = R / (1 + (1-p)R)$, where the $p$ parameter takes into account the optical depth of the torus to its own radiation. An optically thin torus has $p$=0, whereas an optically thick torus has $p$=1 (subset $R_{L13}$). 

    \item As additional checks to make sure that the corrections made to the photometry, as well as the colour selection criterion, have not altered the results, we also present the same results obtained without including these corrections (respectively the subsets ``no-corr'' and ``no-col'').

\end{itemize}

The results evaluated adopting these alternative prescriptions and data sets are shown in Table \ref{tbl:tbl_altcorr}. The structure of the table is the same as that of Table \ref{tbl:tbl_corr_1} and \ref{tbl:tbl_corr_2}.

Although the absolute value of the $CF$ proxy obviously depends on the parameterisation and the wavelength interval for the $L_{\rm NIR}$ and $L_{\rm opt}$ terms, the trends described in the main text are confirmed adopting all the other possible choices. In the data set where the 2--5 $\mu$m interval was adopted a significant correlation appears in between $R$ and $z$ according to the Spearman index. Yet, such trend is not confirmed by neither the Kendall correlation index, nor by the average slope of the fitted regression lines.

\section{The shift of the NIR bump across WISE filters}
\label{app:nir_bump}
The appearance of the NIR bump depends on its sampling in the observed-frame filters. Here we show that, as the redshift increases from $z$=1.0 to $z$=2.9, the NIR bump is sampled only in its rising part, thus leading to the loss of information about its real shape. Here we show this effect at work on the \citet{hernan2016near} template, and how the interpolated/extrapolated SED gets affected.

\begin{figure*}[h!]
\centering
\includegraphics[width=0.8\linewidth,clip]{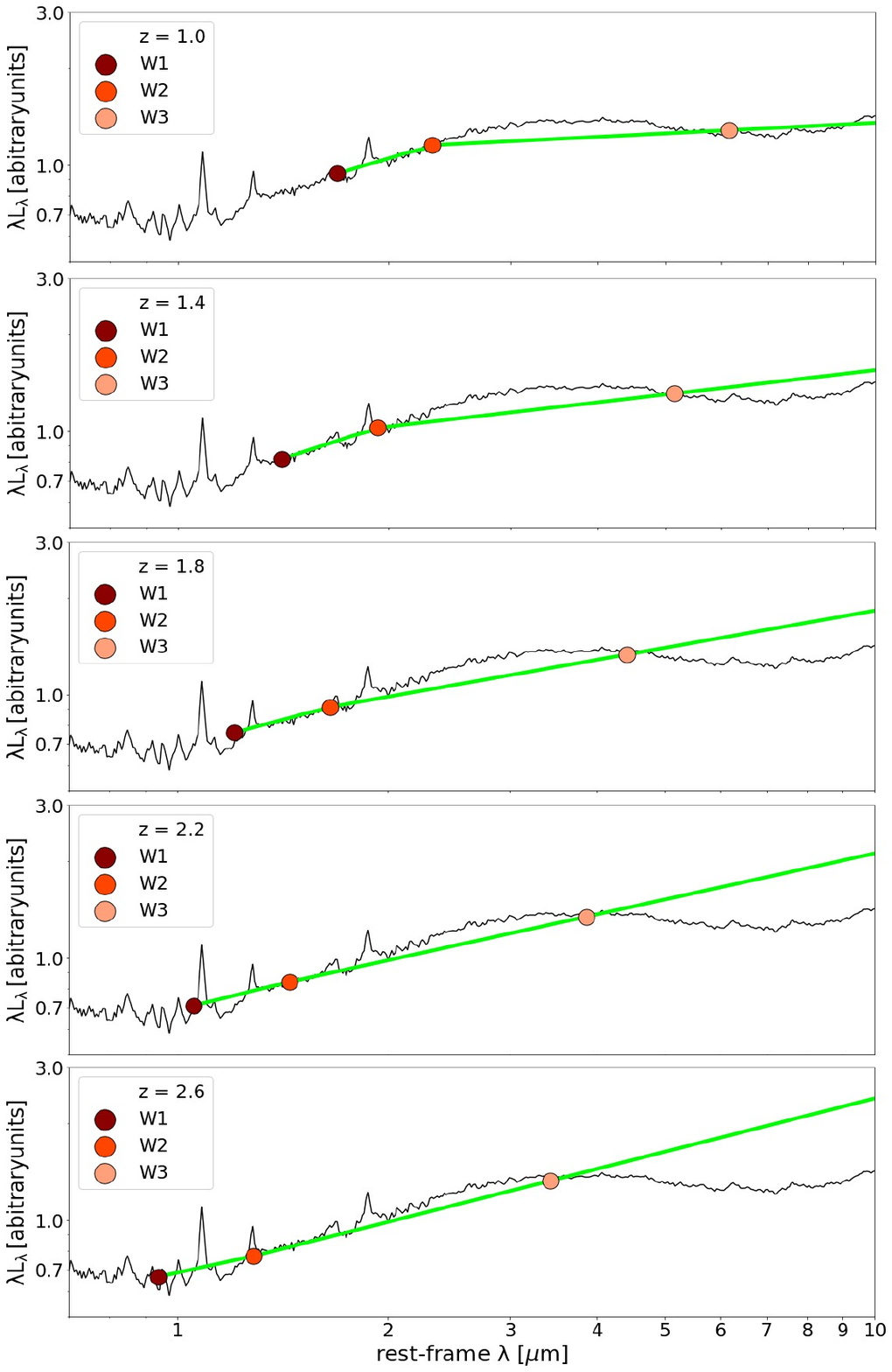}
\caption{The shift of the NIR bump, here represented by the \citet{hernan2016near} template (black), across the WISE W1, W2 and W3 filters for increasing $z$. The green line represents the interpolated/extrapolated SED through the WISE photometric points.}
\label{fig:nir_bump_shift}
\end{figure*}

\end{appendix}

\end{document}